\documentstyle[12pt,epsf]{article}
\newcommand{\nc}{\newcommand}
\nc{\al}{\alpha}
\nc{\g}{\gamma}
\nc{\G}{\Gamma}
\nc{\D}{\Delta}
\nc{\la}{\lambda}
\nc{\La}{\Lambda}
\nc{\sg}{\sigma}
\nc{\var}{\varphi}
\nc{\pa}{\partial}
\nc{\nn}{\nonumber \\ }
\nc{\hf}{\frac{1}{2}}         
\nc{\dz}{\frac{dz}{2\pi i}}
\nc{\binomial}[2]{\left (\begin{array}{c} {#1}\\ {#2} \end{array}
\right )}
\nc{\q}[1]{\lfloor{#1}\rfloor}
\nc{\ben}{\begin{equation}}
\nc{\een}{\end{equation}}
\nc{\bea}{\begin{eqnarray}}
\nc{\eea}{\end{eqnarray}}
\nc{\bra}[1]{\langle {#1}|}
\nc{\ket}[1]{|{#1}\rangle}
\nc{\vj}{V^{(J)}_{-J}}
\nc{\vjpr}{V^{(J')}_{-J'}}
\nc{\smn}{S^{-n}}
\nc{\smnpr}{S^{-n'}}
\newcommand{\Z}{\mbox{$Z\hspace{-2mm}Z$}}
\nc{\C}{\mbox{\hspace{1.24mm}\rule{0.2mm}{2.5mm}\hspace{-2.7mm} C}}
\nc{\Nat}{\mbox{\hspace{.04mm}\rule{0.2mm}{2.8mm}\hspace{-1.5mm} N}}
\newcommand{\R}{\mbox{\hspace{.04mm}\rule{0.2mm}{2.8mm}\hspace{-1.5mm} R}}
\nc{\spa}{\hspace{1 cm},\hspace{1 cm}}
\nc{\vs}{\vspace}
\nc{\NP}[1]{Nucl.\ Phys.\ {\bf #1}}
\nc{\PL}[1]{Phys.\ Lett.\ {\bf #1}}
\nc{\CMP}[1]{Commun.\ Math.\ Phys.\ {\bf #1}}
\nc{\PR}[1]{Phys.\ Rev.\ {\bf #1}}
\nc{\PRL}[1]{Phys.\ Rev.\ Lett.\ {\bf #1}}
\nc{\PTP}[1]{Prog.\ Theor.\ Phys.\ {\bf #1}}
\nc{\PTPS}[1]{Prog.\ Theor.\ Phys.\ Suppl.\ {\bf #1}}
\nc{\MPL}[1]{Mod.\ Phys.\ Lett.\ {\bf #1}}
\nc{\IJMP}[1]{Int.\ Jour.\ Mod.\ Phys.\ {\bf #1}}
\nc{\IM}[1]{Invent.\ Math.\ {\bf #1}}
\nc{\SJNP}[1]{Sov. J. Nucl. Phys.\ {\bf #1}}
\def\beq{\begin{equation}}
\def\eeq{\end{equation}}
\def\hhat{{\widehat h}}
\def\Jhat{{\widehat J}}
\def\Je{J^e{}}
\def\Jehat{{\Jhat^e}{}}
\def\nhat{{\widehat n}}
\def\nonehat{{\widehat n_1}}
\def\ntwohat{{\widehat n_2}}
\def\mhat{{\widehat m}}

\def\qhat{{\widehat q}}
\def\Shat{{\widehat S}}
\def\varpihat{{\widehat \varpi}}
\def\Ub{{\overline U}}
\def\Vb{{\overline V}}
\def\Sb{{\overline  S}}
\def\zb{{\bar z}}
\def\nub{\bar \nu}
\def\varpib{{\overline \varpi}}

\begin{document}

\topmargin -5mm
\oddsidemargin 5mm

\begin{titlepage}
\setcounter{page}{0}

\begin{flushright}
April 1999\\
TIFR/TH/99-18
\end{flushright}

\vs{8mm}
\begin{center}
{\Large Negative Screenings in Conformal Field Theory and 2D Gravity:}\\[.2cm]
{\Large The Braiding Matrix}

\vs{8mm}
{\large J{\o}rgen Rasmussen}\footnote{e-mail address: 
jorgen@theory.tifr.res.in}\\[.2cm]
{\em Department of Theoretical Physics, 
Tata Institute of Fundamental Research}\\
{\em Homi Bhabha Road, Colaba, Mumbai 400 005, India}\\[.2cm] 
and\\[.2cm] 
{\large Jens Schnittger}\footnote{e-mail address: 
schnittg@celfi.phys.univ-tours.fr}\\[.2cm]
{\em Laboratoire de Math\'{e}matiques et Physique Th\'{e}orique}\\
{\em Universit\'{e} de Tours,
Parc de Grandmont, F-37200 Tours, France}

\end{center}

\vs{8mm}
\centerline{{\bf{Abstract}}}
\noindent
We consider an extension of the Coulomb gas picture which is motivated
by Liouville theory and contains negative powers of screening operators
on the same footing as positive ones. 
The braiding problem for chiral vertex operators in this extended
framework is analyzed. We propose  explicit expressions
 for the $R$-matrix with
general integer screening numbers, which are given in terms of 
$_4F_3$ $q$-hypergeometric functions through 
natural  analytic continuations of the well-known expression for positive
integer screenings. These proposals are subsequently verified using a subset of
the Moore-Seiberg equations that is obtained by simple manipulations
in the operator approach.  
 Interesting
new relations for $q$-hypergeometric functions 
(particularly of type $_4F_3$) arise on the way.
\\[.4cm]

\end{titlepage}
\newpage
\renewcommand{\thefootnote}{\arabic{footnote}}
\setcounter{footnote}{0}

\section{Introduction}

The Coulomb gas picture of rational conformal field theory, and its various
generalizations, has amply 
proved to be a very efficient tool for the computation
of conformal blocks, structure constants and braiding properties in these
theories \cite{FF,DF,FFK,PRY}.
Interestingly, Coulomb gas techniques have turned out to describe (certain
sectors of) irrational conformal field theories as well, such as WZNW theory
and Liouville/Toda theory. Within the Gervais-Neveu quantization
of Liouville theory, an efficient formulation of screened vertex operators,
avoiding completely the manipulation of contours, was introduced
long ago in \cite{GN832,Sch90,GS93}. 

The historical development of the exploration of Liouville theory in the 
operator framework was in fact such that for a long time, only primary fields
corresponding to the Kac table were considered, in close analogy 
to the situation
for minimal models. The next step of generalization \cite{GS94,GR94} 
consisted in 
formulating observables with arbitrary conformal dimensions, which however
involved only a {\em non-negative} integer number of screenings. 
It was pointed out in ref. \cite{Sch96} that while those observables 
(the general
Liouville exponentials) are formally described by an infinite sum over 
positive integer powers of screening operators, these infinite sums do not
permit any naive evaluation even in the simple context of three-point 
functions. Depending on the three-point function considered, negative
screening powers can arise as a non-perturbative effect,
establishing contact in this way with the Goulian-Li procedure \cite{GL}.
The analysis of the three-point functions furthermore suggests that
there should exist a conformal algebra, closed under fusion and braiding,
involving chiral vertex operators with both positive and negative integer
powers of screenings. This would constitute a non-trivial generalization of 
the standard Coulomb gas picture and open very interesting perspectives
towards an operator  formulation of a new class of irrational 
conformal field theories.

A great virtue of the Gervais-Neveu approach consists in the fact that
negative integer powers of screenings are just as well defined as positive 
ones so that their introduction does not require any analytic continuation 
procedure. Negative and positive screening powers are in fact 
related by a Weyl 
reflection \cite{BG89,ANPS94}, exchanging one of the two equivalent
free fields
of the Gervais-Neveu approach with  the other. However, except for the cases
considered in \cite{GR94}, so far very little
has been known about the braiding and fusion algebras of vertex operators
involving negative powers of screenings. It is clear that the $R$-matrix
and the fusion matrix should be given by an appropriate analytic continuation
of $q$-deformed $6j$-symbols, as was the case already in the 
generalization from the Kac table to arbitrary continuous spins \cite{GS93}. 

The objective of the present paper is
to establish the braiding algebra of chiral vertex operators
with {\em arbitrary} integer screenings and continuous spins. 
The operator product is then  determined as well by the general
proportionality
relation of Moore and Seiberg \cite{MS} between fusion and braiding matrices.
  
The remaining part of the paper is organized as follows:

In Section 2 we introduce our notation and provide 
some well known background material. 

In Section 3 we present
our proposal (ansatz) for the analytic continuation of the $R$-matrix to 
negative screening numbers. 
The continuation procedure is remarkably simple, and uses no more than
a well-known transformation formula for (truncating) $q$-hypergeometric
sums of type
$_4F_3$, which constitute the essential part of the analytic expression
for the braiding matrix.
Depending on the signs of the screening numbers of the vertex operators
to be braided,  the standard braiding matrix for positive
integer screenings  of refs. \cite{GS93,GS94}  can always be brought
into  a form such that the continuation corresponds to
a simple substitution of  positive by negative screening numbers.

In Section 4 we derive, as a simple consequence
of the polynomial equations of Moore and Seiberg \cite{MS},
a system of determining equations for $R$-matrices with a mix of both
positive
and negative ``ingoing'' screening numbers. We verify that our proposal
fulfills these equations, starting with the case where only one
of the vertex operators to be braided has  negative screening.
In the process we need several new relations for $q$-hypergeometric
functions, in particular of the type $_4F_3$ and generalizations
thereof.
Proofs are outlined and will appear in more detail elsewhere \cite{RS}.
We also discuss the delicate issue of uniqueness of the solution.

In Section 5 we turn to the  class of $R$-matrices where both ingoing
screening numbers are negative. In order to verify our analytic
continuation proposal for this case, we will exploit the fact that it
can be reduced to the ones
treated in Section 4 by means of a concatenation procedure.
Substantial evidence in favour of our proposal is then provided by
checking explicitly
the resulting identity for several non-trivial classes of examples.

In Section 6 we discuss the connection of our extended $R$-matrices
with $6j$-symbols.

Section 7 describes the trivial generalization from only one type of
screening
charge to the case where the conjugate screening charge is included.

Finally, Section 8 is devoted to concluding remarks and a speculative
outlook.
We discuss in particular the non-chiral case in the context of Liouville 
theory, including the strong coupling regime. 

Appendix A  contains some useful observations on certain transformations
of $q$-hyper- \break geometric functions of type ${}_4F_3$, whereas
Appendix B is devoted to further considerations on the uniqueness 
of our proposal
for the braiding matrix.

\section{Coulomb Gas Picture and Liouville Theory}

We start by introducing our notation and by recalling some elementary
facts about the Coulomb gas picture as used in refs. 
\cite{Sch90,GS93,GS94,Sch96}. As usual, Coulomb gas vertex
operators are written as a product of a free field vertex operator and
some power of screening operators. A particularity of the formulation
chosen in the above references is that the integration contour for the 
screenings is fixed once and for all, and does not depend on the 
correlator in question. Concretely, for a vertex operator of spin $J$
and $U(1)$ charge $m$ we have 
\ben
 U_m^{(J)}(\sigma)\equiv V^{(J)}_{-J}(\sigma)S^{J+m}(\sigma)
\een
with
\ben
 V^{(J)}_{-J}(\sigma)=e^{\alpha_- J X}(\sigma), \qquad 
  S(\sigma)=e^{2ih(\varpi+1)}\int_0^\sigma dx V_1^{(-1)}(x)
  +\int_\sigma^{2\pi}dx V_1^{(-1)}(x)
\label{vertexdef}
\een
Here, $X(\sigma)$ is a canonical free field and $\alpha_-$
is the ``semi-classical'' screening charge, related to the central
charge $c$ of the theory in the usual way:
\ben
 c=1+{12\over \alpha_-^2}(1+{\alpha_-^2\over 2})^2
\een
Note that $c$ is arbitrary continuous, and $c>25$ for $\alpha_-$ real. 
In terms of the deformation parameter $h$ ($q=e^{ih}$ is the deformation
parameter relevant for the quantum group interpretation 
\cite{G90,CGR942,CGS97}) we have
\ben
 \alpha_-=\sqrt{2h\over \pi}, \qquad c=1+{6\pi\over h}\left(1+{h
  \over \pi}\right)^2
\een
Moreover, $\varpi$ is essentially $i$ times 
the momentum zero mode of the free field, and is taken to be 
real in the following (this corresponds to the elliptic sector of Liouville
theory \cite{GN82}, and is also the choice appropriate for instance for
the description of minimal models in this framework \cite{G93}).
Explicitly, one has
\ben
 X(\sigma)=q_0+p_0\sigma+i\sum_{n\ne 0} \ e^{-in\sigma}{p_n\over n}
\een 
and $\varpi=ip_0\sqrt{2\pi\over h}$.
The $U_m^{(J)}$ operators are characterized by their conformal weight
$\Delta_J$, their $U(1)$ charge (momentum shift) $m$, and their normalization
$I_m^{(J)}(\varpi)\equiv\bra{\varpi}U^{(J)}_m(\sigma=0)\ket{\varpi+2m}$:
\bea
 \Delta_J&=&-J-{h\over\pi}J(J+1)\nn
 U_m^{(J)}\varpi&=&(\varpi+2m)U_m^{(J)}\nn
 I_m^{(J)}(\varpi)&=&\left( 2 \pi \Gamma(1+{h\over\pi}) \right )^{J+m}
  e^{ ih(J+m)(\varpi-J+m)}\nn
 &\cdot&\prod_{\ell=1}^{J+m}\frac{\Gamma[1+(2J-\ell+1)h/\pi]}{
  \Gamma[1+\ell h/\pi]\Gamma[1-(\varpi+2m-\ell)h/\pi]
  \Gamma[1+(\varpi+\ell)h/\pi]}
\label{3.6}
\eea

The braiding properties and the operator
product of the $U_m^{(J)}$ operators were derived in ref. \cite{CGR941}
in the case of degenerate operators ($2J=0,1,2,...$ and $m=-J,-J+1,...,J$)
and generalized in refs. \cite{GS93,GR94} to the case of arbitrary $J$
and integer positive screening numbers $n=J+m$. Their algebra fulfills
the Moore-Seiberg equations \cite{MS} of conformal field theory \cite{GR94}. 

For later use, we introduce a special notation for the leading order operator
product (fusion) of the two operators $U_m^{(J)}$ and $U_{m'}^{(J')}$:
\ben
 U^{(J)}_m(\sigma)\odot U^{(J')}_{m'}(\sigma):= \lim_{\sigma'\to\sigma}
  \frac{U^{(J)}_m(\sigma)\cdot U^{(J')}_{m'}(\sigma')}{
  (1-e^{i(\sigma'-\sigma)})^{-\Delta_J-\Delta_{J'}+\Delta_{J+J'}}}
\label{odotdef}
\een
(as usual, $\sigma'$ has to be given a small positive imaginary part
to make the above expression well defined). One has the simple multiplication
law
\ben
 U_{m}^{(J)}(\sigma)\odot U_{m'}^{(J')}(\sigma)=q^{2J'(J+m)}U_{m+m'}^{(J+J')}
  (\sigma)
\label{odot}
\een
for any (positive or negative) integer values of the screenings. 

Local observables (in the Liouville context they are the Liouville 
exponentials)
can be constructed as bilinear combinations of the $U_m^{(J)}$ and their
right moving counterparts ${\overline U}_m^{(J)}$, with $\varpi$-dependent
coefficients \cite{Sch90,G93,GS93}. Locality is a simple
consequence of the orthogonality relations obeyed by the $q$-deformed
$6j$-symbols that constitute the essential part of the chiral braiding
matrices. In the present paper, we shall concentrate on the algebra
of the chiral vertex operators; the application of the extended formalism
with negative screenings to the construction of local observables, 
as considered in \cite{Sch96},
is left for a future publication. 

A crucial feature of the Gervais-Neveu approach is the introduction of
a second free field, related to the one above by a quantum 
canonical transformation \cite{GN832}. As pointed out in ref. \cite{Sch96},
this is of direct relevance for the problem of constructing negative (integer)
powers of screening operators: Negative powers of screening operators 
in terms of $X$
are nothing else than positive powers of screening operators in terms of
$\tilde X$, the second free field. Therefore, there is no need of any
analytic continuation procedure to define negative screenings, 
and the corresponding
ambiguities are absent from the start. 
Concretely, we have (cf. ref. \cite{Sch96})
\ben
 S^{-1}(\sigma)=\tilde S(\sigma)\frac{1}{I^{({\hf})}_{\hf}(\varpi+1)
  I^{({\hf})}_{\hf}(-\varpi-1)}
\label{ss}
\een
where $\tilde S$ is the screening operator constructed from $\tilde X$.

We remark here that, $X$ and $\tilde X$ being completely 
equivalent, the replacement of one by the other must not change any physical
observable. {}From the quantum group point of view, this means that observables
are invariant not only under infinitesimal $U_q(sl(2))$ transformations, but 
also under Weyl reflections \cite{BG89,ANPS94}. As discussed in
\cite{Sch96}, for Liouville exponentials outside the Kac table 
($2J\ne 0,1,2,...$),
this is a highly non-trivial condition which is in conflict with naive
charge conservation rules familiar from the standard Coulomb gas picture. It
serves, in fact, as an important guideline for the correct {\em 
non-perturbative}
evaluation of matrix elements of the Liouville exponentials as constructed
within the Gervais-Neveu approach. 

\section{The Braiding Matrix}

The braiding matrix $R$ describes the braiding of the two chiral fields
$U_{m}^{(J)}(\sigma)$ and $U_{m'}^{(J')}(\sigma')$ 
\ben
 U_{m}^{(J)}(\sigma)U_{m'}^{(J')}(\sigma')=\sum_{n_1,n_2}
  R(J,J';\varpi)_{n,n'}^{n_2,n_1}U_{m_2}^{(J')}(\sigma')U_{m_1}^{(J)}(\sigma)
\label{Rdef}
\een
The ordering of $\sigma$ and $\sigma'$ is implicit in this definition of
$R$ and we shall 
deal with the case $0<\sigma<\sigma'<2\pi$ explicitly, reserving the
notation $\overline{R}$ for the opposite ordering of $\sigma$ and $\sigma'$.
The sums extend over the integers
\ben
 n_1=J+m_1\spa n_2=J'+m_2
\een
the ranges of which we shall discuss below. 
The parameters are subject to the condition
\ben
 m_1+m_2=m+m'
\een
so that there is really just one sum in Eq. (\ref{Rdef}).
In general, the combinations 
\ben
 n=J+m\spa n'=J'+m'
\een
are related to the screening numbers and have mainly been treated 
as {\em non-negative} integers in the literature. 
In particular, the braiding matrix for non-negative integers $n$ and $n'$
and arbitrary spins is known to be \cite{GS93} 
\bea
 &&R(J,J';\varpi)_{n,n'}^{n_2,n_1}=\exp\left\{-i\pi(\D_x+\D_{x+m+m'}
  -\D_{x+m_2}-\D_{x+m})\right\}q^{2nJ'-2n_2J}\nn
 &\cdot&\q{2x-2J'+2n_2+1}\binomial{n}{n_1}\nn
 &\cdot&\frac{\q{2x+n_2+2}_{n_1}
 \q{2x-2J-2J'+n+n_2+1}_{n'}\q{2J'-n_2+1}_{n-n_1}}{\q{2x-2J'+n_2+1}_{n+n'+1}}\nn
 &\cdot&{}_4F_3\left(\begin{array}{l}-2J+n,\ -2J'+n_2,\ -n_1,\ -n'\\ 
    -2x-n-n'-1,\ n-n_1+1,\ 2x-2J-2J'+n+n_2+1 \end{array};\ q,\ 1
  \right)\nn
 &=&\exp\left\{-i\pi(\D_x+\D_{x+m+m'}
  -\D_{x+m_2}-\D_{x+m})\right\}q^{2nJ'-2n_2J}\nn
 &\cdot&\q{2x-2J'+2n_2+1}\binomial{n}{n_1}\nn
 &\cdot&\frac{\q{2x+n_2+2}_{n_1}
 \q{2x-2J-2J'+n+n_2+1}_{n'}\q{2J'-n_2+1}_{n-n_1}}{\q{2x-2J'+n_2+1}_{n+n'+1}}\nn
 &\cdot&\sum_{l=0}^{\infty}\frac{\q{-2J+n}_l\q{-2J'+n_2}_l\q{-n_1}_l\q{-n'}_l}{
  \q{l}!\q{-2x-n-n'-1}_l\q{n-n_1+1}_l\q{2x-2J-2J'+n+n_2+1}_l}
\label{R}
\eea
with $x$ defined by $\varpi=2x+1+{\pi\over h}$. 
Here, the standard notation for the $q$-hypergeometric function has been
used
\bea
 {}_4F_3\left(\begin{array}{llll}a,&b,&c,&d\\  e,&f,&g&{} \end{array}
 ;\ q,\ \rho
  \right)&=&\sum_{n=0}^\infty\frac{\q{a}_n\q{b}_n\q{c}_n\q{d}_n}{
   \q{e}_n\q{f}_n\q{g}_n\q{n}!}\rho^n\nn
 \q{a}_0=1\spa\q{a}_n&=&\q{a}\q{a+1}...\q{a+n-1}
\label{F}
\eea
It should be noted that in the standard case ($n$ and $n'$ non-negative 
integers) the summation in (\ref{R}) truncates after a {\em finite} number of 
terms. We shall use the convention $\q{y}=\sin(yh)/\sin(h)$.

A $q$-hypergeometric function (\ref{F}) is balanced if $g=a+b+c+d-e-f+1$
and Saalschutzian if furthermore $d$ (or equivalently $a$, $b$ or $c$)
is a non-positive integer. A $q$-deformed Saalschutzian ${}_4F_3$
hypergeometric function satisfies the standard transformation (ST)
rule \cite{GR94}
\bea
 {}_4F_3\left(\begin{array}{l}a,\ b,\ c,\ d\\ 
    e,\ f,\ g\end{array};\ q,\ 1\right)
 &=&\frac{\q{f-c}_{-d}\q{e+f-a-b}_{-d}}{\q{f}_{-d}\q{e+f-a-b-c}_{-d}}\nn
 &\cdot&{}_4F_3\left(\begin{array}{l}e-a,\ e-b,\ c,\ d\\ 
    e,\ e+f-a-b,\ c+d-f+1 \end{array};\ q,\ 1\right)
\label{ST}
\eea
The present work is devoted to a discussion
of the braiding matrix for the case where the ingoing screening
numbers are {\em arbitrary integers}. We will assume that the 
outgoing screening numbers are also
integers - see below for further comments on this point.

As the braiding matrix for non-negative integer screenings Eq. (\ref{R})
is expressed in terms of a $q$-hypergeometric function, the most
naive procedure for an extension to negative screenings would be to simply
replace one or both of the ingoing screening numbers $n,n'$ by negative
integers and simultaneously allow the outgoing screening numbers to be
arbitrary integers. 
However, this would in general lead to $q$-hypergeometric sums that do
not truncate, and such sums may well diverge for $|q|=1$ since
the denominators of individual terms cannot be bounded away from zero. 
The remedy is given by employing ST (\ref{ST}) {\em before} 
continuing the ingoing screening numbers after which a continuation
results in a {\em finite} and well defined expression, as will be discussed
below. We thus arrive at an
explicit proposal for the $R$-matrix, which will subsequently be verified.

\subsection{Analytic Continuation}

{}From a mathematical point of view, the extension of 
the explicit expression (\ref{R}) to arbitrary integer screenings
is certainly not unique
as we are trying to analytically continue a function given only on a discrete
set of points. Let us make a division into subcases characterized by the
signs of the ingoing screening numbers and treat them one by one.
The expressions for the $R$-matrices we
obtain should of course only be considered as an ansatz. The evidence
for their validity will be given in the following sections. 

\subsubsection{{\bf I}: Positive-Positive Case, $n,n'\geq0$}

It is well known \cite{GS93} that the set of chiral vertex operators
corresponding to non-negative screening numbers closes under braiding,
ensuring that both of the outgoing operators have non-negative 
screening numbers.
It is easily verified algebraically that precisely then is the $R$-matrix 
(\ref{R}) non-vanishing and 
has no poles in the limits where $2J$ or $2J'$ becomes
an integer\footnote{Such an analysis is similar
in spirit  to the discussion of fusion rules in ref. \cite{PRY}.}. 
The finite summation range for the truncating
$q$-hypergeometric function is given explicitly by
\ben
 max(0,n_1-n)\leq l\leq min(n_1,n')
\label{selI}
\een
and $n_1,n_2\geq0$. This analysis suggests that a more
natural expression exists for $n_1>n$,  which allows to avoid
 cancellations of the
form $\q{-1}!/\q{-1}!$. Let us introduce the notation
$R_>^I$ and $R_\leq^I$ for the type {\bf I} $R$-matrix ($n,n'\geq0$)
in the cases
$n_1>n$ and $n_1\leq n$, respectively. By construction, $R_\leq^I$
is given by (\ref{R}) whereas $R_>^I$ is naturally represented as
\bea
 &&R_>^I(J,J';\varpi)_{n,n'}^{n_2,n_1}=\exp\left\{-i\pi(\D_x+\D_{x+m+m'}
  -\D_{x+m_2}-\D_{x+m})\right\}q^{2nJ'-2n_2J}\nn
 &\cdot&\q{2x-2J'+2n_2+1}\binomial{n'}{n_1-n}\nn
 &\cdot&\frac{\q{2x+n_2+2}_n\q{2x-2J-2J'+n+n'+1}_{n_2}\q{-2J+n}_{n_1-n}}{
  \q{2x-2J'+n_2+1}_{n+n'+1}}\nn
 &\cdot&{}_4F_3\left(\begin{array}{l}-2J+n+n'-n_2,\ -2J'+n',\ -n,\ -n_2\\ 
    2x-2J-2J'+n+n'+1,\ -2x-n-n_2-1,\ n'-n_2+1 \end{array};\ q,\ 1\right)
\label{RI>}
\eea
This expression may be obtained either by shifting the summation over
$n_1-n\leq l\leq n'$ to $0\leq l\leq n_2$, or by a repeated use of ST
(\ref{ST}). In Appendix A some useful techniques based on ST are discussed
and the present situation is presented as an illustration.

\subsubsection{{\bf II}: Negative-Positive Case, $n<0\leq n'$}

Inspection of Eqs. (\ref{R}) and (\ref{RI>}) shows that
$R^{II}$ may be represented by exactly the
same mathematical expressions as $R^I$ when in the latter $n$ has been
continued to negative integers:
\ben
 R^{II}_\leq=R^I_\leq\spa R^{II}_>=R^I_>
\label{II=I}
\een
For the second equality to make sense, it is crucial that
Eq. (\ref{RI>})
vanishes term by term for $n_2<0$ due to the binomial prefactor. 
$R^{II}_\leq$ also vanishes in this case.
An alternative way to obtain Eq. (\ref{II=I}) is the following:
In order to have a well defined and non-vanishing $R$-matrix for $n<0\leq n'$
and $2J$ and $2J'$ non-integer, we find that the 
summation variable $l$ in Eq. (\ref{R}) must satisfy
\ben
 0\leq l\leq min(n_1-n-1,n')\ \ {\mbox{or}} \ \ max(0,n_1-n)\leq l\leq n'
\een 
It should be noted that there is no overlap between these two regimes.
The first regime is excluded by our second demand that the $R$-matrix
must be well defined in the limit where $2J'$ is an integer (analyzing
the limit where $2J$ is an integer does not provide new information).
This leaves us with 
\ben
 max(0,n_1-n)\leq l\leq n'\ \ \rightarrow\ \ 0\leq n_2
\label{selII}
\een
The last inequality expresses a selection rule preserving the
non-negativity of the screening number under braiding (from the left)
with a negative screening vertex operator.

\subsubsection{{\bf III}: Positive-Negative Case, $n'<0\leq n$}

In complete analogy with the analysis of case {\bf II} we find that the type
{\bf III} $R$-matrix may be represented by the following 
single surviving range for the summation variable $l$ (\ref{R}) 
\ben
 max(0,n_1-n)\leq l\leq n_1\ \ \rightarrow\ \ 0\leq n_1
\label{selIII}
\een
Accumulating the information from cases {\bf I}, {\bf II} and {\bf III},
we observe that the property of non-negativity of a screening number 
is preserved under any braiding; see also Section 4.

\subsubsection{{\bf IV}: Negative-Negative Case, $n,n'<0$}

In the final case where both $n$ and $n'$ are negative integers our
construction goes as follows. We shall initially focus on the
$q$-hypergeometric part of $R^I$ which will be denoted $I$.
Employing ST once results in the rewriting
\bea
 I&=&{}_4F_3\left(\begin{array}{l}-2J+n,\ -2J'+n_2,\ -n-n'+n_2,\ -n'\\ 
    -2x-n-n'-1,\ -n'+n_2+1,\ 2x-2J-2J'+n+n_2+1 \end{array};\ q,\ 1\right)\nn
 &=&\frac{\q{-2x+2J'-n-n'-n_2-1}_{n'}\q{-2x+2J-n-n'}_{n'}}{\q{-2x-n-n'-1}_{n'}
  \q{-2x+2J+2J'-n-n'-n_2}_{n'}}\nn
 &\cdot&{}_4F_3\left(\begin{array}{l}2J-n-n'+n_2+1,\ n+1,\ 
  -2J'+n_2,\ -n'\\ 
    -n'+n_2+1,\ -2x+2J-n-n',\ 2x-2J'+n+n_2+2 \end{array};\ q,\ 1\right)
\label{Ianal}
\eea
It should be stressed that a possible pole or zero in $I$ which is 
matched by an appropriate zero or pole, respectively, in the prefactors
of $R^I$, must also be present in the right hand side. In this sense, 
the two expressions are completely equivalent.
Now, the right hand side is also well defined after the substitution $n_i
\rightarrow-n_i-1$. This is the core of our analytic continuation and
we {\em define} the object $I_{-n-1.-n'-1}^{-n_2-1,-n_1-1}$ by the right
hand side in (\ref{Ianal}) after the substitution. The full type {\bf IV}
$R$-matrix is obtained by multiplying $I_{-n-1.-n'-1}^{-n_2-1,-n_1-1}$ 
by the original prefactors in $R^I$ in which the substitution 
$n_i\rightarrow -n_i-1$ has been performed.
The shift by $-1$ in the
substitution $n_i\rightarrow -n_i-1$ is convenient because
$R^{IV}$ is defined only for purely negative ingoing screening numbers.
Nevertheless, let us state our proposal in the form
\bea
 &&R^{IV}(J,J';\varpi)_{-n,-n'}^{-n_2,-n_1}
  =\exp\left\{-i\pi(\D_x+\D_{x+m+m'}
  -\D_{x+m_2}-\D_{x+m})\right\}q^{-2nJ'+2n_2J}\nn
 &\cdot&\q{2x-2J'-2n_2+1}\binomial{-n}{-n_1}\nn
 &\cdot&\frac{\q{2x-n_2+2}_{n_2-1}
  \q{2x-2J-2J'-2n-n'-n_2+2}_{n-1}\q{2J'+n_2+1}_{n'-n_2}}{
  \q{2x-2J-2n-n'+2}_{n+n'-1}}\nn
 &\cdot&{}_4F_3\left(\begin{array}{l}-2J-n,\ -2J'-n_2,\ 1-n_2,\ 1-n\\ 
    -2x,\ n'-n_2+1,\ 2x-2J-2J'-2n-n'-n_2+2 \end{array};\ q,\ 1
  \right)
\label{RIV}
\eea
where an additional ST has been employed. Here $n$ and $n'$ are {\em positive}
integers and $-n_i=J_i+m_i$.
\\[.2cm]
In conclusion, our two analytic continuation procedures result
in the same expressions for the braiding matrices. In case {\bf IV}, though,
only one approach seems to apply directly. In \cite{RS}
we shall return to the question of infinite sum representations
of $q$-hypergeometric functions and braiding matrices.

We have seen that all four types may be constructed
using well defined concatenations of ST followed by trivial
continuations of one or two of the ingoing screening numbers.
Thus, the approach based on repeated use of ST demonstrates the 
{\em universality} of the mathematical expression for the
original braiding matrix. Of course, this statement
pertains to our proposal only. In the following sections we shall
argue for the validity of this proposal. 

\subsection{Weyl Reflection Symmetry}

In this section we shall perform a first test on our proposal by comparing it 
with the braiding matrices obtainable by Weyl reflection symmetry from the
well known type {\bf I}. Since Weyl reflections act on the screening
numbers as $n_i\rightarrow 2J_i-n_i$, subclasses of all four types are
reachable, depending on the signs of $2J-n$ and $2J'-n'$. 
As we are not addressing the question of non-integer screening
numbers we shall restrict ourselves to the case $2J,2J'\in\Z$. 

The explicit relation between $\tilde{S}^n$ and $S^{-n}$ is
\bea
 \tilde{S}^n&=&S^{-n}K_n(\varpi)\nn
 K_n(\varpi)&=&K_1(\varpi+2(n-1))K_1(\varpi+2(n-2))...K_1(\varpi)
\eea
where $K_1(\varpi)$ is given by the normalization factor in the right hand
side of Eq. (\ref{ss}) 
\ben
 K_1(\varpi)=I^{(\hf)}_\hf(\varpi+1)I^{(\hf)}_\hf(-\varpi-1)
\een
Weyl reflection dictates that
\bea
 R(J,J';\varpi)^{-k_2,-k_1}_{-n,-n'}&=&
  R(J,J';-\varpi)^{k_2,k_1}_{n,n'}
  \frac{I^{(J)}_J(\varpi)I^{(J')}_{J'}(\varpi+
  2J-2n)}{I^{(J')}_{J'}(\varpi)I^{(J)}_J(\varpi+2J'-2k_2)}\nn
 &\cdot&\frac{K^{(J')}_{k_2}(\varpi+2J'-2k_2)K_{k_1}(\varpi+2J+2J'-2k_2-2k_1)
  }{K_n(\varpi+2J-2n)K_{n'}(\varpi+2J+2J'-2n-2n')}
\eea
and one may show that this reduces to
\ben
 q^{2nJ'-2n_2J}R(J,J';-\varpi)_{2J-n,2J'-n'}^{2J'-n_2,2J-n_1}
  =q^{-2nJ'+2n_2J}R(J,J';\varpi)_{n,n'}^{n_2,n_1}
\label{ref}
\een
This relation may be seen as boundary conditions on our proposal imposed
by the reflection symmetry. 

Due to the universality of the {\em mathematical}
structure of the braiding matrix, it is almost straightforward
to verify that our proposal satisfies (\ref{ref}). 
In principle, one should distinct between the four
resulting types of braiding matrices. However, we know from Appendix A
that Weyl reflections for $2J,2J'\in\Z$ and (concatenations of) 
the STs we employ commute so it is sufficient to consider only one type.
In the representation of type {\bf I} we have
\bea
 &&q^{2nJ'-2n_2J}R^I(J,J';-\varpi)_{2J-n,2J'-n'}^{2J'-n_2,2J-n_1}\nn
 &=&\exp\left\{-i\pi(\D_x+\D_{x+m+m'}
  -\D_{x+m_2}-\D_{x+m})\right\}\q{2x-2J'+2n_2+1}\binomial{2J-n}{2J-n_1}\nn
 &\cdot&\frac{\q{2x-2J-2J'+n+n'+1}_{2J-n_1}
  \q{2x-2J'+n+n'+n_2+2}_{2J'-n'}\q{n_2+1}_{n'-n_2}}{
  \q{2x-2J-2J'+n+n'+n_2+1}_{2J+2J'-n-n'+1}}\nn
 &\cdot&{}_4F_3\left(\begin{array}{l}-2J+n_1,\ -2J'+n',\ -n_2,\ -n\\ 
    -2x-n-n_2-1,\ n'-n_2+1,\ 2x-2J-2J'+n+n'+1 \end{array};\ q,\ 1\right)\nn
 &=&q^{-2nJ'+2n_2J}R^I(J,J';\varpi)_{n,n'}^{n_2,n_1}
\eea
where the last equality is due to a concatenation of STs.

\section{Double Braiding Relations for the Type II and III $R$-matrices}

In this section we shall set up simple relations for the $R$-matrices
using one of the polynomial equations of Moore and Seiberg \cite{MS},
namely the commutativity of fusion and braiding. 
The Moore-Seiberg equations were originally set up within the context of
rational conformal field theory, but can be viewed as a set of
consistency relations
for conformal field theory in general. For the present operator algebra,
which extends the standard Coulomb gas in a rather non-trivial fashion,
their validity may not seem self-evident -  especially since both
fusion
and braiding matrices involve an infinite number of conformal blocks.
We will take the commutativity of fusion and braiding as an
axiomatic starting point of our analysis, though in principle
our explicit operatorial construction should allow to discuss its
validity.
In any such discussion, manipulations with infinite sums over conformal
operators
will necessarily arise and one may speculate that the Moore-Seiberg
equations
will act as a defining principle for their treatment; clearly, further
analysis will
be necessary to elucidate this point.

There are several possibilities for choosing an appropriate equation
system to study. It turns out that a convenient choice
is the recursive system to be analyzed in the following.

\subsection{Recursive Equation Systems}

Let us first consider the product
\ben
 U_{-m_1}^{(-J_1)}(\sigma)\odot
U_{m_2}^{(J_2)}(\sigma)U_{m_3}^{(J_3)}(\sigma')
\een
(cf. Eq. (\ref{odotdef}))
and braid through from the right the operator
$U_{m_3}^{(J_3)}(\sigma')$.
Demanding that the above expression can be evaluated either by
braiding first and then fusing the operators at $\sigma$, or vice
versa,  allows us to derive the
double braiding relation
\bea
 &&\sum_{l=0}^{n_2+n_3} \sum_k \, q^{2J_2(n_1-k)}
R(J_2,J_3;\varpi-2m_1)_{n_2,n_3}^{n_2+n_3-l,l}
  R(-J_1,J_3;\varpi)_{-n_1,n_2+n_3-l}^{-n_1+n_2+n_3-l+k,-k}\nn
&\cdot&U_{m_3+n_2-n_1-l+k}^{(J_3)}(\sigma')U_{-(J_2-J_1)+l-k}^{(-J_1+J_2)}
(\sigma)
\nn
 &=&\sum_iR(-J_1+J_2,J_3;\varpi)_{-n_1+n_2,n_3}^{-n_1+n_2+n_3-i,i}
  U_{n_2-n_1+m_3-i}^{(J_3)}(\sigma')U_{i-(J_2-J_1)}^{(-J_1+J_2)}(\sigma)
\label{n1n2n3}
\eea
$n_1$ is a positive integer while $n_2$ and $n_3$ are non-negative ones.
The sums over $k$ and $i$ could in principle contain non-integer values
or even an integration, since they are related to outgoing screening
numbers about
which we have no a priori knowledge as soon as we leave the safe grounds
of the standard Coulomb gas with positive screenings only.
In Appendix B, we present an argument involving degenerate field
techniques which suggests that non-integer values may
actually occur, with the corresponding $R$-matrix elements  decoupling
from those for integer outgoing screenings. This is quite puzzling, as
one would expect our explicit operator construction to select one
particular $R$-matrix, removing the ambiguity. However, braiding
problems with negative
ingoing screenings involve infinite sums over products of outgoing vertex
operators, the evaluation of which is expected to be delicate
\cite{Sch96}; in particular, several equivalent representations with
numerically different
$R$-matrices may exist, as is suggested by the argument in Appendix B.
These questions, though clearly  important, go beyond the scope of this
first analysis of the braiding problem with negative screenings. We
will therefore
discuss here only the $R$-matrix elements for integer outgoing
screenings.
We consider the case $n_3=0$ in Eq. (\ref{n1n2n3}) for which we may
deduce
the equation system\\[.2cm]
{\bf Triangular Equation System}
\bea
 &&\sum_{l=0}^{n_2}q^{2J_2(n_1+L-l)}
  R(J_2,J_3;\varpi-2m_1)_{n_2,0}^{n_2-l,l}
  R(-J_1,J_3;\varpi)_{-n_1,n_2-l}^{-n_1+n_2-L,L-l}\nn
 &=&R(-J_1+J_2,J_3;\varpi)_{-n_1+n_2,0}^{-n_1+n_2-L,L}
\label{tri}
\eea
by comparison of coefficients of like operators. For integer
outgoing screenings, $L$ is integer.  
Note that in any case there is no coupling between matrix elements for
integer and non-integer outgoing screenings.
Due to the ``triangular'' structure of this equation system, all braiding 
matrices of type {\bf II} are determined recursively in terms 
of $R$-matrices of type {\bf I}
and the subclass of type {\bf II} consisting of $R$-matrices of the form
$R(J,J';\varpi)_{-n,0}^{n_2,-n_1}$. However, the latter may be determined
by an additional and independent equation system and a selection
rule. 
Consider
the product
\ben
 U_{n/2}^{(n/2)}(\sigma)\odot S^{-n}(\sigma)V_{-J'}^{(J')}(\sigma')
\een
and braid through from the right the operator $V_{-J'}^{(J')}(\sigma')$. 
One obtains  the double braiding relation
\bea
 &&\sum_{n_1,n_1'}R(0,J';\varpi+n)_{-n,0}^{n_1-n,-n_1}
  R(n/2,J';\varpi)_{n,n_1-n}^{n_1-n-n_1',n_1'}\nn
 &\cdot&V_{-J'}^{(J')}(\sigma')
  S^{n_1-n-n_1'}(\sigma')V_{-n/2}^{(n/2)}(\sigma)S^{n_1'-n_1}(\sigma)\nn
 &=&e^{ihnJ'}V_{-J'}^{(J')}(\sigma')V_{-n/2}^{(n/2)}(\sigma)
\eea
Now we may use that degenerate fields remain degenerate fields after
braiding \cite{GN84} and we derive the equation system\\[.2cm]
{\bf Initial Equation System}
\ben
 e^{-ihnJ'}\sum_{l=0}^n \, R(n/2,J';\varpi)_{n,K+l-n}^{K,l}
  R(0,J';\varpi+n)_{-n,0}^{K+l-n,l-K}=\delta_{K,0}
\label{ini1}
\een
which we shall denote the ``initial'' one. This equation system 
fixes the remaining indeterminacy, if we impose the additional
selection rule\\[.2cm]
{\bf Selection Rule}
\bea
 R(J_1,J_2;\varpi)_{-n, n'}^{n_2,n'-n-n_2}=0 
  \ \ \ \ \mbox{for}\ \ n>0,\ n'\geq0,\ n_2<0\nn
 R(J_1,J_2;\varpi)_{n,-n'}^{n-n'-n_1,n_1}=0
  \ \ \ \ \mbox{for}\ \ n\geq0,\ n'>0,\ n_1<0
\label{selrule}
\eea
All screening numbers are taken to be
integers. The selection rule means that positive screenings always remain
positive screenings after braiding. This rule is certainly fulfilled
by our analytic continuation of the $R$-matrix as already discussed, 
but we do not have, at present, an a priori derivation. Its validity
can be verified in the case where one of the ingoing
vertex operators is degenerate, cf. Appendix B. It is an interesting
question whether it can be deduced in general by invoking other
polynomial equations, or by using more information about the explicit 
structure of the negative screening vertex operators. 

If we consider the case of integer $K$ and take into account the
selection rule, we can rewrite Eq. (\ref{ini1}) as 
\ben
 e^{-ihnJ'}\sum_{l=max(0,K-n)}^KR(n/2,J';\varpi)_{n,l}^{K,n+l-K}
  R(0,J';\varpi+n)_{-n,0}^{l,-n-l}=\delta_{K,0}
\label{ini}
\een
This equation system now determines recursively all remaining
unknowns, if we consider successively $K=0,1,2...$.
We will now outline the proof that our explicit proposal
is the unique solution of  Triangular and Initial  equation systems
plus selection rule, if only integer outgoing screenings are admitted.
The argument is based on certain new
identities for $q$-hypergeometric functions and generalizations
thereof. Detailed proofs of these identities
will be presented elsewhere \cite{RS}.

\subsection{Verification of Proposal}

Here we shall verify that the analytically continued expression
(\ref{R}) subject to (\ref{selII}) provides a solution to the two double 
braiding relations discussed above. 

We first consider the triangular equation system (\ref{tri}).
Insertion of our proposal yields the equation system
\bea
 &&\sum_{l=0}^{n_2}\q{2x+2J_1-2J_3-2n_1+2n_2-2l+1}\binomial{n_2}{l}
  \tilde{R}(-J_1,J_3;\varpi)_{-n_1,n_2-l}^{-n_1+n_2-L,-l+L}\nn
 &\cdot&\frac{\q{2x+2J_1-2n_1+n_2-l+2}_l\q{2J_3-n_2+l+1}_{n_2-l}}{
  \q{2x+2J_1-2J_3-2n_1+n_2-l+1}_{n_2+1}}\nn
 &=&\Theta(-n_1+n_2-L\geq0)\q{2x-2J_3-2n_1+2n_2-2L+1}(-1)^{L}
  \frac{\q{n_1-n_2}_{L}}{\q{L}!}\nn
 &\cdot&\frac{\q{2x-n_1+n_2-L+2}_L\q{2J_3+n_1-n_2+L+1}_{-n_1+n_2-L}}{
  \q{2x-2J_3-n_1+n_2-L+1}_{-n_1+n_2+1}}
\label{n1n20}
\eea
where
\bea
 &&\tilde{R}(-J_1,J_3;\varpi)_{-n_1,n_2-l}^{-n_1+n_2-L,-l+L}\nn
 &=&
  \q{2x-2J_3-2n_1+2n_2-2L+1}(-1)^{L-l}\frac{\q{n_1}_{L-l}}{\q{L-l}!}
  \q{2x-n_1+n_2-L+2}_{L-l}\nn
 &\cdot&\frac{\q{2x+2J_1-2J_3-2n_1+n_2-L+1}_{n_2-l}
  \q{2J_3+n_1-n_2+L+1}_{-n_1-L+l}}{\q{2x-2J_3-n_1+n_2-L+1}_{-n_1+n_2-l+1}}\nn
 &\cdot&\sum_{i=max(0,n_1+L-l)}^{n_2-l}\frac{\q{2J_1-n_1}_{i}\q{-2J_3-n_1
  +n_2-L}_{i}}{\q{i}!\q{-2x+n_1-n_2+l-1}_{i}}\nn
  &\cdot&\frac{\q{-L+l}_{i}\q{l-n_2}_i}{\q{-n_1-L+l+1}_{i}
   \q{2x+2J_1-2J_3-2n_1+n_2-L+1}_{i}}
\label{hat}
\eea
$\tilde{R}$ is merely the $R$-matrix (\ref{R}) subject to
(\ref{selII}) without the explicit phases and powers of $q$
which cancel out their analogues in the other $R$-matrices.
Our proof will be by induction in $n_2$ and it is recalled that 
$n_1\geq1$. Introduce the integer
\ben
 N=-n_1+n_2-L
\een
Due to the selection rule, the equation system becomes trivial
for $N<0$, so in the following we may assume that $N\geq0$.

It is easily verified that the Triangular equation system (\ref{n1n20})
is satisfied for the initial value $n_2=0$.
For general $n_2$, the left hand side in (\ref{n1n20}) will be denoted 
${\cal L}$ whereas the right hand side will be denoted ${\cal R}$.
The finite double summation in ${\cal L}$ may be expressed as
\ben
 {\cal L}=\sum_{j\geq0}S_j
\label{sumS}
\een
where
\bea
 &&S_j\nn
 &=&(-1)^{n_2}\q{2x-2J_3+2N+1}\frac{\q{n_2}!\q{n_1+N-j-1}!
  \q{2J_3-n_2-N+j+1}_{n_2+N-j}}{\q{n_2-j}!
  \q{n_1-1}!\q{j}!\q{N-j}!\q{-2x-N-1}_{n_1+N-j}}\nn
 &\cdot&\frac{\q{-2J_1+n_1-n_2+j+1}_{n_2-j}}{
  \q{-2x-2J_1+2J_3+2n_1-2n_2-1}_{n_2+1}
  \q{-2x+2J_3+n_1-n_2-N-1}_{-n_1+n_2+1}}\nn
 &\cdot&\sum_{l=0}^{n_2-j}\q{2x+2J_1-2J_3-2n_1+2n_2-2l+1}
  \q{-2x-2J_1+2J_3+n_1-n_2-N+l}_j\nn
 &\cdot&\frac{\q{-n_2+j}_l
  \q{-2x-2J_1+2n_1-n_2-1}_l}{\q{l}!\q{-2J_1+n_1-n_2+j+1}_l}\nn
 &\cdot&\frac{\q{-2x-2J_1+2J_3+2n_1-2n_2-1}_l\q{-2x+2J_3+n_1-n_2-N-1}_l}{
  \q{2J_3-n_2-N+j+1}_l\q{-2x-2J_1+2J_3+2n_1-n_2}_l}\nn
\eea
We also introduce
\ben
 S_j(\left\{-p\right\})=S_j(2x+p,2J_1,2J_3-p,n_1,n_2-p,N-p)
\label{shiftS}
\een
where the parameters $n_2,\ N,\ -2x$ and $2J_3$ all have been subtracted
$p$ while the remaining parameters $n_1$ and $2J_1$ are left unchanged.
${\cal R}(\left\{-p\right\})$ is defined analogously where the unshifted
${\cal R}$ is given explicitly by
\ben
 {\cal R}=-\frac{\q{2x-2J_3+2N+1}\q{n_1-n_2+N-1}!\q{2J_3-N+1}_N
  \q{-2x+2J_3-N}_{n_1-n_2-1}}{\q{N}!\q{n_1-n_2-1}!\q{-2x-N-1}_{n_1-n_2+N}}
\label{calR}
\een

We shall make a case distinction and consider $N=0$ first in which case
\ben
 {\cal L}=S_0
\een
{\bf Lemma 1}
\ben 
 S_0={\cal R}(2x,2J_1,2J_3,n_1,n_2,N=0)
\label{1}
\een
{}\\[.2cm]
For $N>0$, let $M_k$ denote the following sum
\bea
 M_k&=&(-1)^{n_2}\q{2x-2J_3+2N+1}\frac{\q{n_1+N-1}!
  }{\q{n_1-1}!\q{k}!\q{N}\q{N-k-1}!}\nn
 &\cdot&\frac{\q{2J_3-n_2-N+k+1}_{n_2+N-k}\q{-2J_1+n_1-n_2+k+1}_{n_2-k}}{
  \q{-2x-N-1}_{n_1+N-k}\q{-2x-2J_1+2J_3+2n_1-2n_2+k-1}_{n_2-k+1}}\nn
 &\cdot&\frac{1}{
  \q{-2x+2J_3+n_1-n_2-N-1}_{-n_1+n_2+1}}\nn
 &\cdot&\sum_{l=0}^{n_2-k}\q{2x+2J_1-2J_3-2n_1+2n_2-2l+1-k}\nn
 &\cdot&\frac{
  \q{-n_2+k}_l\q{-2x-2J_1+2n_1-n_2-1}_l}{\q{l}!\q{-2J_1+n_1-n_2+k+1}_l}\nn
 &\cdot&\frac{\q{-2x-2J_1+2J_3+2n_1-2n_2+k-1}_l\q{-2x+2J_3+n_1-n_2-N-1}_l}{
  \q{2J_3-n_2-N+k+1}_l\q{-2x-2J_1+2J_3+2n_1-n_2}_l}
\label{Mk}
\eea
where $M_0=S_0$.
The following lemma determines a relation between $S_j$ and $M_k.$\\[.2cm]
{\bf Lemma 2}\\[.2cm]
For $k<n_2$ we have
\ben
 M_k-M_{k+1}=\sum_{m=0}^{k+1}A_k^mS_m(\left\{-k-1+m\right\})
\label{Mk'}
\een
where
\ben
 A_k^m=(-1)^{k+m}\frac{\q{n_1-n_2+N-1}_{k+1-m}
  \q{2J_3-k+m}_{k+1-m}}{\q{k+1-m}!\q{-2x+2J_3-N-k-1+m}_{k+1-m}}
\label{A}
\een
and in particular $A_k^{k+1}=-1$.\\[.2cm]
It then follows that
\ben
 {\cal L}=\sum_{k=0}^{min(N,n_2)-1}
  \sum_{m=0}^kA_k^mS_m(\left\{-k-1+m\right\})+\Theta(n_2<N)M_{n_2}
\een
from which (\ref{n1n20})
may be deduced, thus completing the proof of the Triangular equation system 
(\ref{tri}).

Now we turn to the Initial equation system (\ref{ini}).
Insertion of our proposal yields the equation system
\bea 
 &&\delta_{K,0}\nn
 &=&\q{2x-2J'+2K+1}(-1)^n\frac{\q{-n}_{n-K}\q{2x+K+2}_{n-K}
  \q{-2J'}_K\q{2x-2J'+2}_{n-1}}{\q{n-K}!\q{2x-2J'+K+1}_{n+1}\q{2x+2}_n}\nn
 &\cdot&\sum_{l=max(0,K-n)}^K
  \q{2x-2J'+n+2l+1}\frac{\q{-K}_l\q{2x-2J'+K+1}_l}{\q{l}!\q{n-K+1}_l}\nn
 &\cdot&\frac{\q{n}_l\q{2x-2J'+n+1}_l}{\q{2x-2J'+n+K+2}_l\q{2x-2J'+2}_l}
\label{ini2}
\eea
It is immediate that the right hand side vanishes for $K<0$ and that
it reduces to 1 for $K=0$. We may therefore assume $K>0$. After a case
distinction into $0<K\leq n$ and $0\leq n<K$ the identity follows by
induction in $K$ and $n$, respectively, thus completing the proof
of the Initial equation system (\ref{ini}). Details on the proofs of both
the Triangular and the Initial equation systems will appear 
elsewhere \cite{RS}.

\subsection{The Type {\bf III} $R$-matrix}

So far we have only verified our proposal for $R$-matrices
of type {\bf II}. Type {\bf III} $R$-matrices may be obtained 
from the latter  essentially by hermitian conjugation. We have
\bea
 \left( V^{(J)}_{-J}(\sg)S^n(\sigma)V^{(J')}_{-J'}(\sg')
  S^{-n'}(\sigma') \right)^\dagger
 &=&S^{-n'\dagger}(\sg')V^{(J')\dagger}_{-J'}(\sigma') 
  S^{n\dagger}(\sg)V^{(J)\dagger}_{-J}(\sigma)\nn
 &=&\tilde S^{-n'}(\sg')\tilde V^{(J')}_{-J'}(\sigma')\tilde S^n(\sg)
  \tilde V^{(J)}_{-J}(\sigma)
\eea
Here we have used that for real $\varpi$, the two free fields are just
hermitian conjugates of each other \cite{GN832} 
($\tilde V^{(J')}_{-J'}$ and $\tilde S$ 
denote operators constructed from the field $\tilde X$). 
The order of screening operators and free field exponentials can be reverted
to the standard form at the expense of simple phase factors, so that
\ben
 \left( V^{(J)}_{-J}(\sg)S^n(\sigma)V^{(J')}_{-J'}(\sg')
  S^{-n'}(\sigma') \right)^\dagger
 =q^{2(Jn-J'n')}\tilde V^{(J')}_{-J'}(\sg')\tilde S^{-n'}(\sigma')
  \tilde V^{(J)}_{-J}(\sg)\tilde S^n(\sigma)
\label{conj1}
\een
The expression on the right hand side is of the type {\bf II}
already discussed. Braiding the operators on the left hand side with
the unknown type {\bf III} $R$-matrix, we obtain 
\ben
 R^*(J,J';\varpi-2n_1+2n_2+2(J+J'))_{n,-n'}^{-n_2,n_1}\ q^{2(Jn_1-J'n_2)}
  =q^{2(Jn-J'n')}{\overline R} (J',J;-\varpi)_{-n',n}^{n_1,-n_2}
\een
The bar on the $R$-matrix on the right hand side indicates that this
$R$-matrix braids an operator at $\sigma'$ with an operator at $\sigma$
(cf. comment following Eq. (\ref{Rdef})),
contrary to the one on the left hand side. Notice also the replacement
$\varpi \to -\varpi$ in this $R$-matrix which reflects the fact that
it applies to the braiding of ``tilded'' operators. 
The final result for the type {\bf III} braiding matrix thus becomes
\ben
 R(J,J';\varpi)_{n,-n'}^{-n_2,n_1}=q^{-2[J(n-n_1)-J'(n'-n_2)]}
  {\overline R}^*(J',J; -\varpi-2n_1+2n_2+2(J+J'))_{-n',n}^{n_1,-n_2}
\label{conj2}
\een

In the case of braiding of positive screening vertex operators, it
is well known that ${\overline R}(J,J';\varpi)_{n,n'}^{n_2,n_1}$
differs from $R(J,J';\varpi)_{n,n'}^{n_2,n_1}$ only by a phase factor
\cite{CGR941},
and it is evident from our analytic continuation procedure
that the type {\bf II} $R$-matrix will keep this property.
Therefore, we can write Eq. (\ref{conj2}) as a relation between $R$-matrices
corresponding to the same position ordering:
\bea
 R(J,J';\varpi)_{n,-n'}^{-n_2,n_1}&=&q^{-2[J(n-n_1)-J'(n'-n_2)]} 
  e^{-2i\pi(\Delta_x +\Delta_{x+m+{m}'}
  -\Delta_{x+{m}_2}-\Delta_{x+m})}\nn
 &\cdot&R^*(J',J; -\varpi-2n_1+2n_2+2(J+J'))_{-n',n}^{n_1,-n_2}
\label{conj3}
\eea
where ${m}'=-n'-J'$ and ${m}_2=-n_2-J'$. Eq. (\ref{conj3}) determines 
the type {\bf III} $R$-matrix in terms of the type  {\bf II} one,
which we have already verified. We show that our analytic continuation
fulfills Eq. (\ref{conj3}) by considering separately the two cases
$n\geq n_1$ and $n<n_1$.  From our formula for the type {\bf II}
$R$-matrix, we obtain  for the 
right hand side (up to explicit phases
and powers of $q$, cf. comment following Eq. (\ref{hat}))
\bea
 &&\tilde{R}(J',J;-\varpi-2n_1+2n_2+2(J+J'))_{-n',n}^{n_1,-n_2}=\q{-2x+2J'
  +2n_2-1}\binomial{-n'}{-n_2}\nn
 &\cdot&\frac{\q{-2x+2J+2J'-n_1+2n_2}_{-n_2}\q{-2x-n+n_2-1}_n
  \q{2J-n_1+1}_{n_2-n'}}{\q{-2x+2J'-n_1+2n_2-1}_{n-n'+1}}\nn
 &\cdot&\sum_{l=max(0,n'-n_2)}^n\frac{\q{-2J'-n'}_l\q{-2J+n_1}_l
  \q{n_2}_l\q{-n}_l}{\q{l}!\q{2x-2J-2J'+n_1-n_2+}_l\q{n_2-n'+1}_l
  \q{-2x-n+n_2}_l}
\label{Rstar}
\eea
In the first case $n\geq n_1$, the summation in (\ref{Rstar}) may be written
\bea
 &&\sum_{l=n'-n_2}^n\ \rightarrow\ \frac{\q{-n}_{n'-n_2}\q{n_2}_{n'-n_2}}{
  \q{n'-n_2}!\q{n_2-n'+1}_{n'-n_2}}\nn
 &\cdot&\frac{\q{-2J'-n'}_{n'-n_2}\q{-2J+n_1}_{n'-n_2}}{\q{2x-2J-2J'+n_1
  -n_2+1}_{n'-n_2}\q{-2x-n+n_2-1}_{n'-n_2}}\nn
 &\cdot&\sum_{l=0}^{n_1}\frac{\q{-2J'-n_2}_l\q{-2J+n}_l\q{n'}_l\q{-n_1}_l}{
  \q{l}!\q{n'-n_2+1}_l\q{2x-2J-2J'+n-n_2+1}_l\q{-2x-n+n'-1}_l}
\eea
and by re-inserting it we recognize the left hand side of Eq. (\ref{conj3}).
In the second case $n<n_1$, the summation on the left 
hand side of Eq. (\ref{conj3}) may be written
\bea
 &&\sum_{l=n_1-n}^{n_1}\
  \rightarrow\ \frac{\q{-n_1}_{n_1-n}\q{n'}_{n_1-n}}{\q{n_1-n}!
  \q{n-n_1+1}_{n_1-n}}\nn
 &\cdot&\frac{\q{-2J+n}_{n_1-n}\q{-2J'-n_2}_{n_1-n}}{\q{-2x-n+n'-1}_{n_1-n}
  \q{2x-2J-2J'+n-n_2+1}_{n_1-n}}\nn
 &\cdot&\sum_{l=0}^n\frac{\q{-2J+n_1}_l\q{-2J'-n'}_l\q{-n}_l\q{n_2}_l}{\q{l}!
  \q{n_1-n+1}_l\q{-2x+n_2-n-1}_l\q{2x-2J-2J'+n_1-n_2+1}_l}
\eea
and by re-inserting it we recognize the expression (\ref{Rstar}). 
This concludes the verification of our proposal for the type {\bf III}
$R$-matrix.

\section{Connection with $6j$-symbols}

The braiding matrix for vertex operators from the Kac table was shown in
\cite{CGR941} to coincide with $q$-deformed $6j$-symbols for
$U_q(sl(2))$,
up to normalization factors that were determined explicitly. This result
continues to be true for general positive screening vertex operators
\cite{GS93,GR94}. We will not show here that our expression for
the $R$-matrix in the presence of negative screenings can indeed
be interpreted consistently as a $q$-deformed $6j$-symbol - this would
require verifying all of the defining properties of the latter -
but rather offer a most natural generalization of the standard 
relation between $R$-matrix and $6j$-symbol, as given in ref. \cite{GS94}:
\ben
 R(J,J',\varpi)_{n,n'}^{n_2,n_1}=
  e^{-i\pi (\Delta_c+\Delta_b-\Delta_e-\Delta_f)}
  {\kappa_{ab}^e \kappa_{de}^c\over \kappa_{db}^f \kappa_{af}^c}
  \left\{ ^{a}_{d}\,^{b}_{c}\right. \left |^{e}_{f}\right\}
\label{6j}
\een
The arguments of the $q$-deformed $6j$-symbol are
\bea
 a=J,\quad b=x+m+m',\quad c=x\nn
  d=J',\quad e=x+m_2,\quad f=x+m
\label{3.21}
\eea
and the coefficients $\kappa_{J_1 J_2}^{J_{12}}$ are given by
\bea
 \kappa_{J_1 J_2}^{J_{12}}&=&
  \left ({ he^{-i(h+\pi)}
  \over 2\pi \Gamma(1+h/\pi) \sin h}\right )^{J_1+J_2-J_{12}}
  e^{ih (J_1+J_2-J_{12})( J_1-J_2-J_{12})} \nn
 &\cdot&
  \prod_{k=1}^{J_1+J_2-J_{12}}
  \sqrt{ \lfloor 1+2J_1-k\rfloor \over
  \lfloor k \rfloor \, \lfloor 1+2J_2-k \rfloor\,
  \lfloor -(1+2J_{12}+k) \rfloor }
\label{3.9}
\eea
The  last equation  makes sense
for arbitrary $J_1,J_2, J_{12}$ such that $J_1+J_2-J_{12}$ is a
non-negative integer, but is readily extended to negative
screening numbers if we define products with negative upper limits
as usual by
\beq
\prod_{j=1}^{-n}f(j):={1\over \prod_{j=1}^n f(1-j)}
\label{negprod}
\eeq
As our method provides us directly with the braiding matrix, rather than
the $6j$-symbol, Eq. (\ref{6j}) should be read as our {\em definition}
of (or proposal for) the $q$-deformed $6j$-symbol, 
extended to arbitrary integer screening numbers.

An extension to the purely negative case where all (ingoing 
and outgoing)
screening numbers are negative, has been provided in ref. \cite{GR94}.
Here we shall show that our definition (\ref{6j}) based on our proposal
for the braiding matrix, reproduces (and generalizes\footnote{In
  contrast to ref. \cite{GR94}, we do not assume that the
  outgoing screenings are negative.})  the
result of \cite{GR94} in the purely negative case. The idea is to verify
invariance of our proposal under the replacement $J_i\rightarrow-J_i-1$ 
(implying in particular that screening numbers are transformed
according to $n_i\rightarrow-n_i-1$)
\ben
 \left\{ ^{J}_{J'}\,^{x-J-J'+n+n'}_{x}\right. 
  \left |^{x-J'+n_2}_{x-J+n}\right\}=
 \left\{ ^{-J-1}_{-J-1'}\,^{-x+J+J'-n-n'-1}_{-x-1}
  \right. \left |^{-x+J'-n_2-1}_{-x+J-n-1}\right\}
\label{GR}
\een
which is the defining property for the extension
in ref. \cite{GR94}. In order to do that
let us write down explicitly the well known $q$-deformed $6j$-symbol
\bea
 &&\left\{ ^{J}_{J'}\,^{x-J-J'+n+n'}_{x}
  \right. \left |^{x-J'+n_2}_{x-J+n}\right\}\nn
 &=&\left(\frac{\q{2J-n+1}_{n_2-n'}\q{2x-2J-2J'+n+n'+n_2+1}_{n-n_2}}{
  \q{2J'-n_2+1}_{n_2-n'}\q{2x+n_2+2}_{n-n_2}}\right.\nn
 &\cdot&\left.\frac{\q{2x-2J'+n_2+1}_{n+n'+1}\q{2x-2J+2n+1}\q{n_2}!\q{n_1}!}{
  \q{2x-2J+n+1}_{n+n'+1}\q{2x-2J'+2n_2+1}\q{n}!\q{n'}!}\right)^{\hf}\nn
 &\cdot&R^I(J,J';\varpi)_{n,n'}^{n_2,n_1}
\label{6jexp}
\eea
and our proposal in the purely negative case
\bea
 &&\left\{ ^{-J-1}_{-J-1'}\,^{-x+J+J'-n-n'-1}_{-x-1}
  \right. \left |^{-x+J'-n_2-1}_{-x+J-n-1}\right\}\nn
 &=&\left(\frac{\q{-2J+n}_{n'-n_2}\q{-2x+2J+2J'-n-n'-n_2}_{n_2-n}}{
  \q{-2J'+n_2}_{n'-n_2}\q{-2x-n_2-1}_{n_2-n}}\right.\nn
 &\cdot&\left.\frac{\q{-2x+2J'-n_2}_{-n-n'-1}\q{-2x+2J-2n-1}\q{-n_1-1}!
  \q{-n_2-1}!}{
  \q{-2x+2J-n}_{-n-n'-1}\q{-2x+2J'-2n_2-1}\q{-n-1}!\q{-n'-1}!}\right)^{\hf}\nn
 &\cdot&R^{IV}(-J-1,-J'-1;-\varpi)_{-n-1,-n'-1}^{-n_2-1,-n_1-1}
\eea
Employing ST six times, it is now straightforward to verify
the invariance (\ref{GR}).

In ref. \cite{GR94} it is observed that the $q$-deformed $6j$-symbol
possesses the simple symmetry
\ben
 \left\{ ^{J'}_{J}\,^{x-J-J'+n+n'}_{x}
  \right. \left |^{x-J+n}_{x-J'+n_2}\right\}
 =\left\{ ^{J}_{J'}\,^{x-J-J'+n+n'}_{x}
  \right. \left |^{x-J'+n_2}_{x-J+n}\right\}
\label{6jsym}
\een
An explicit proof based on (\ref{6jexp}) is immediate, so let us
conclude this section by stating the equivalence of (\ref{6jsym})
in terms of $R$-matrices
\bea
 &&q^{2nJ'-2n_2J}R^I(J',J;\varpi)_{n_2,n_1}^{n,n'}\nn
 &=&\frac{\q{2x-2J+2n+1}\q{n_2}!\q{n_1}!\q{2x-2J-2J'+n+n'+n_2+1}_{n-n_2}}{
  \q{2x-2J'+2n_2+1}\q{n}!\q{n'}!\q{2x+n_2+2}_{n-n_2}}\nn
 &\cdot&\frac{\q{2J-n+1}_{n_2-n'}\q{2x-2J'+n_2+1}_{n+n'+1}}{
  \q{2J'-n_2+1}_{n_2-n'}\q{2x-2J+n+1}_{n+n'+1}}
  q^{-2nJ'+2n_2J}R^I(J,J';\varpi)_{n,n'}^{n_2,n_1}
\eea

\section{Construction of the Type {\bf IV} $R$-matrix by Concatenation}

In this section we shall argue for the validity of our proposal for the 
type {\bf IV} $R$-matrix by explicit comparison with an expression 
obtained by a 3 step concatenation. 

Step 1 is a direct verification that our proposal for
$J=J'=0$ is identical to the result of imposing Weyl reflection symmetry
on the corresponding type {\bf I} $R$-matrix. Eq. (\ref{ref}) 
readily ensures that. Step 2 is then the generalization to $J'=0$ and 
$J$ generic, using the result of Step 1. Finally, Step 3 completes 
the generalization to $J$ and $J'$ generic.
In both Step 2 and Step 3, we consider the braiding of 
$\vj\smn(\sigma)$ with $\vjpr\smnpr(\sigma')$, for $n,n'>0$.

\subsection{Step 2}

The $R$-matrix we want to establish is given by
\ben
 \vj(\sg)\smn(\sg)\smnpr(\sg')=\sum_{n_2}R(J,0;\varpi)^{-n_2,-n_1}_{-n,-n'}
 S^{-n_2}(\sg')\vj(\sg)S^{-n_1}(\sg)
\een
where $n_1\equiv n+n'-n_2$.
On the other hand we also have ($k_2\equiv n+n'-k_1$)
\ben
 \vj(\sg)\smn(\sg)\smnpr(\sg')=\vj(\sg)\sum_{k_1} \,
  R(0,0;\varpi)^{-k_2,-k_1}_{-n,-n'}S^{-k_2}(\sg')S^{-k_1}(\sg)
\een
where $0\le k_1,k_2\le n+n'$ by reflection symmetry (whereby the sum
over $k_1$ is seen to be
finite). The right hand side can be further rewritten as ($l_1\equiv
l_2-k_2$)
\ben
 \sum_{k_1}\, R(0,0;\varpi-2J)^{-k_2,-k_1}_{-n,-n'} \  \sum_{l_2} \,
R(J,0;\varpi)
  ^{-l_2,l_1}_{0,-k_2} \ S^{-l_2}(\sg')\vj(\sg)S^{l_1-k_1}(\sigma)
\een
leading to the equation\\[.2cm]
{\bf Step 2 Equation}
\ben
 R(J,0;\varpi)^{-n_2,n_2-n-n'}_{-n,-n'}=\sum_{k=0}^{min(n_2,n+n')} \, 
  R(0,0;\varpi-2J)^{-k,k-n-n'}_{-n,-n'} \ 
  R(J,0;\varpi)^{-n_2,n_2-k}_{0,-k}
\label{step2}
\een
The right hand side is fully known as the type {\bf III} braiding has been
solved. The upper summation limit $n_2$ is due to the selection rule
and could also be inferred by a ``fusion rule analysis'' (see Section
3.1) of the left hand side.
Though we do not have a complete proof of (\ref{step2}) we do have
substantial evidence since we have been able to prove it for $n_2=0,1$ or 2. 

\subsection{Step 3}

It is firstly noted that 
\ben
 \vj(\sg)\smn(\sg)\vjpr(\sg')\smnpr(\sg')=q^{2J'n'}
  \vj(\sg)\smn(\sg)\smnpr(\sg')\vjpr(\sg')
\een
We will consider this last expression,
assuming that our proposal for $R(J,0;\varpi)^{-n_2,n_2-n-n'}_{-n,-n'}$
has already been verified (Step 2). We then have
\bea
 &&\vj(\sg)\smn(\sg)\smnpr(\sg')\vjpr(\sg')\nn
 &=&\sum_{k_2\ge 0} \, R(J,0;\varpi)^{-k_2,k_2-n-n'}_{-n,-n'}\,
  S^{-k_2}(\sg')\, \sum_{l_2}\, R(J,J';\varpi)^{l_2,l_1}_{k_2-n-n',0}\nn
 &\cdot&\vjpr(\sg')S^{l_2}(\sg')\vj(\sg)S^{l_1}(\sg)\nn
 &=&\sum_{k_2\ge 0}\,q^{-2J'k_2} R(J,0;\varpi)^{-k_2,k_2-n-n'}_{-n,-n'}
  \ \sum_{l_2}\, R(J,J';\varpi-2k_2)^{l_2,l_1}_{k_2-n-n',0}\nn
 &\cdot&\vjpr(\sg')S^{l_2-k_2}(\sg')\vj(\sg)S^{l_1}(\sg)
\eea
in addition to
\ben
 \vj(\sg)\smn(\sg)\vjpr(\sg')\smnpr(\sg')=
 \sum_{n_2} R(J,J';\varpi)^{-n_2,-n_1}_{-n,-n'}
\vjpr(\sg')S^{-n_2}(\sg')
  \vj(\sg)S^{-n_1}(\sg)
\een
with $l_1,n_1$ again defined by conservation of total screening number.
By comparing the two results for the braiding in question, we find\\[.2cm]
{\bf Step 3 Equation}
\ben
 R(J,J';\varpi)^{-n_2,n_2-n-n'}_{-n,-n'}= \sum_{k\ge 0} \, q^{2J'(n'-k)}
  \, R(J,0;\varpi)^{-k,k-n-n'}_{-n,-n'}
  R(J,J';\varpi-2k)^{k-n_2,n_2-n-n'}_{k-n-n',0}
\label{step3}
\een
By analyzing the second factor on the right hand side we find that the
summation over $k$ is restricted as 
\bea
 n_2\leq k&\mbox{for}&n+n'\leq n_2\nn
 n_2\leq k\leq n+n'-1&\mbox{for}&n_2\leq n+n'-1
\eea
An explicit proof of Eq. (\ref{step3}) is quite laborious, but we have
checked it explicitly in the case $n=1$ and $n_2\le n'$, thus presenting
further evidence in favour of our proposal.

A further discussion on the type {\bf IV} $R$-matrix will appear 
elsewhere \cite{RS}.

\section{Case of Two Screening Charges: $\alpha_+$ and $\alpha_-$}

Up to now, we have restricted ourselves to vertex operators involving
only screenings corresponding to the ``semi-classical'' screening charge
$\alpha_-$. However, it is straightforward to include the screenings
corresponding to the conjugate charge $\alpha_+$, with $\alpha_+\alpha_-=
2\pi$. 
The corresponding deformation parameter is $\hhat \equiv {\pi^2\over h}$.
Denoting by $\Shat$ the screening operators corresponding to
$\alpha_+$, one may consider the general vertex operators 
$U_{m^e}^{(J^e)}$ given by
\beq
 U_{m^e}^{(J^e )}:=V^{(\Je)}_{-\Je}S^{n}{\hat S}^{\nhat}
\label{doublecharge}
\eeq
with $n,\nhat$ integer and $m^e= n+\nhat {\alpha_+\over\alpha_-}-J^e$.
One may then follow exactly the same reasoning as
in ref. \cite{GS94} to obtain the braiding matrix for $U_{m\mhat}^{(J\Jhat)}$
from that of the single screening vertex operators; the essential observation
here is that screening operators of opposite types commute, $[S(\sigma),
\Shat(\sigma')]=0$.\footnote{It is quite remarkable that the technique
used in refs. \cite{GS93,GS94} to derive the braiding matrix
continues to work in the presence of
negative screenings. We have been able to show 
that the fundamental equations (2.17)
of ref. \cite{GS93} and (5.7) of ref. \cite{GS94} continue to hold, when
using the standard continuation (\ref{negprod})
to products with negative upper limits. These equations could in fact
have been used as an alternative starting point for our analysis.}
One thus obtains, exactly as in the positive screening case,
\beq
 R(\Je,{\Je}';\varpi)_{
\underline{n},\underline{n}'}^{\underline{n}_2,
\underline{n}_1}=q^{-\Je{\Je}'}\qhat^{-\Jehat{\Jehat}'}
R(\Je,{\Je}';\varpi )_{n,n'}^{n_2,n_1}
\hat R(\Jehat,{\Jehat}';\varpihat )_{\nhat, {\nhat}'}
^{\ntwohat, \nonehat}
\label{doubleR}
\eeq
The conventions here are the following: $\underline{n}=(n,\nhat)$ represents
the two screening numbers corresponding to $\alpha_-$ and $\alpha_+$, 
$\Jehat =\Je{h\over\pi}=\Je{\alpha_-\over\alpha_+}$,  
$\varpihat=\varpi{h\over\pi}$ and $\qhat\equiv e^{i\hhat}$. Finally,
$\hat R$ is the same function of its arguments as $R$, except that
$\alpha_-$ is replaced by $\alpha_+$ everywhere (or equivalently
$h$ by $\hhat$). 

\section{Conclusions and Outlook}

What have we learnt from the present analysis? First of all, we
believe to have presented convincing evidence that there is a closed exchange
algebra for a set of generalized Coulomb gas vertex operators involving
positive and negative powers of screenings alike, and we have obtained
the explicit form of the braiding matrix in this general context, through
natural analytic continuations of the braiding matrix for 
positive screenings. The basic lesson
is that the additional structure generated by the inclusion of negative
screenings is essentially determined by the algebra for vertex operators
with positive screenings alone, through fundamental consistency conditions
of conformal field theory. This is because negative screening operators are,
in a rather precise sense, inverses of positive screening operators.
Our results can be expressed, as in the positive screening case, 
in terms of $_4F_3$ $q$-hypergeometric functions which truncate. 

As pointed out in Appendix B, the question of uniqueness of our solution
is more delicate than might have been naively expected. Whether there
is a deeper meaning to the formal possibility of introducing, in our
analysis,  $R$-matrix elements
for non-integer outgoing screenings merits further investigation.
Likewise, the origin of the selection rule of Section 4.1 should be
elucidated. The technique employed in ref. \cite{GS93} 
(cf. footnote in Section 7), which uses
detailed input from the operatorial construction of the vertex operators,
may allow to shed some light on both of these questions.

To complete the description of the conformal algebra, we also need the
fusion matrix. While it should be given in general in terms of the
braiding matrix through one of the Moore-Seiberg relations, one could also 
think of an independent derivation along similar lines as for the
braiding matrix. It appears that
this is in fact possible, and one obtains recursion relations from
the associativity (${\cal F}{\cal F}{\cal F}={\cal F}{\cal F}$) relation.
We hope to come back to this question in a future publication.  

\begin{figure}
\epsfxsize=10truecm
\epsfysize=10truecm
\epsffile{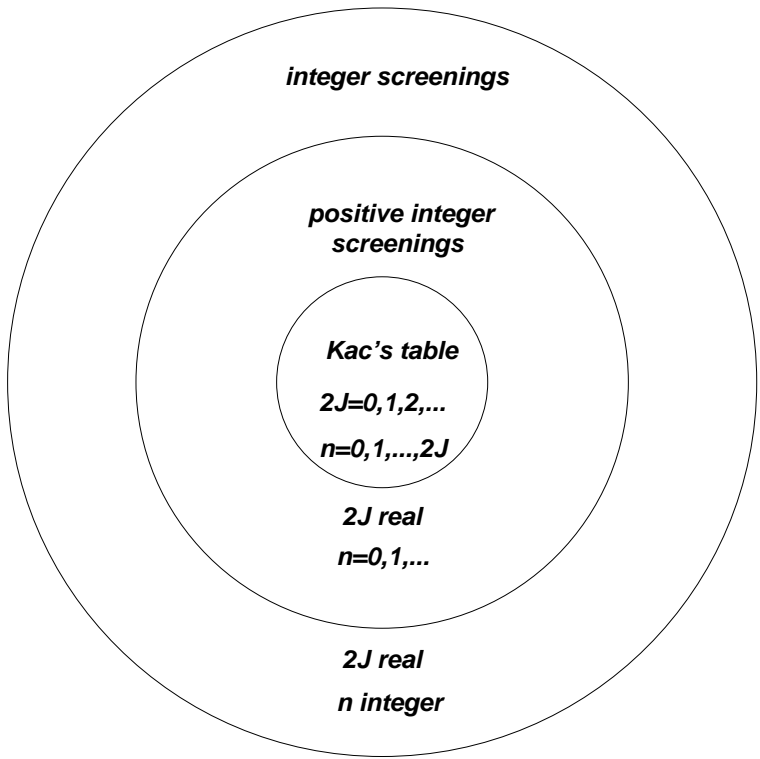}
\caption[]{``Shell model'' of chiral vertex operator algebra}
\label{shellmodel}
\end{figure}

The growing body of results thus
indicates strongly that on the level of the chiral
operator algebra, there exists a sequence of inclusions\footnote{See
ref. \cite{GR94} for a discussion on how the Kac table is embedded
into the larger shells from the point of view of the polynomial
equations.}, each representing
a consistent solution of the Moore-Seiberg equations; see Fig. 
\ref{shellmodel}.
The situation will, however, be much more complicated on the non-chiral 
level, as will be discussed below.

Another interesting direction for future investigations  would be an
analysis of the quantum group aspect of the generalized Coulomb gas 
picture advocated here. It is well known \cite{B88,G90,CGR942} that there
exists another basis of chiral vertex operators $\xi^{(J)}_M$ 
which renders the
underlying $U_q(sl(2))$ symmetry of the algebra manifest; braiding and
fusion symbols for these vertex operators are given by  universal
$R$-matrix and $3j$-symbols for $U_q(sl(2))$, respectively. The ``covariant''
vertex operators are related to those of the Coulomb gas picture by
a basis transformation with $\varpi$-dependent coefficients $|J,\varpi)_M^m$. 
In succession of our work ref. \cite{Sch96}, we were able to
carry out formally
an analysis similar to the one above within the quantum group
covariant basis, with the role of the screening number $n$
played by $N=J+M$. It is not hard to obtain the braiding matrix for 
the case where one or both of $N,N'$ are negative (integer). 
 The analysis is in fact
much simpler than in the Coulomb gas basis as the algebraic expression
for the $R$-matrix in the covariant basis is essentially just a ratio
of $q$-factorials, and the generalized $R$-matrix can essentially
be obtained by naive analytic continuation. Similar selection rules
as for the Coulomb gas basis hold. However, the covariant vertex
operators for generic $2J$ are given by infinite sums 
over Coulomb gas vertex operators,  the precise meaning of which is not 
a priori clear. In particular, by replacing one free field by the other,
one would seem to obtain two  different copies of the same
quantum group representation. 
It is interesting to speculate whether non-perturbative mechanisms along
the lines of ref. \cite{Sch96} will render these two copies equivalent. 
We hope to return to the study of these questions in the future.

\subsection{Non-chiral Case and Liouville Theory}

Our analysis so far has addressed the chiral operator algebra only; let us
now make some more speculative remarks on the conformal field theories that
could be obtained by combining the two chiralities, and in particular
about the relevance of our analysis for Liouville theory. The general rule for 
constructing the non-chiral theory is that operators of the latter should
have crossing-symmetric correlators, and obey a consistent
closed fusion and braiding 
algebra. To find the general answer to this question is, obviously, a 
formidable task. However, 
within the operator approach to Liouville theory, two mechanisms
for producing local fields have been identified. In the standard ``weak
coupling'' ($c>25$) regime,  locality arises from the orthogonality
relations for the $q$-deformed $6j$-symbols. In fact, one writes the local
fields - the Liouville exponentials - as diagonal
combinations of left and right moving vertex operators, both in the sense
of conformal weight and $U(1)$ quantum number $m$:
\beq
e^{-J\alpha_-\Phi}=\sum_{n\ge 0} a_n^{(J)}(\varpi) U_m^{(J)}(u)\Ub_m^{(J)}(v)
\label{Liou}
\eeq
with $u:=\tau+\sigma, \ v:=\tau-\sigma$. The coefficients $a_n^{(J)}(\varpi)$
can actually be absorbed in the normalization of the vertex operators, upon 
which the braiding of the latter will be given in terms of $q$-deformed
$6j$-symbols
alone \cite{GS94}. Mutual locality of two Liouville exponentials 
then becomes the statement that the contraction
of left and right moving $R$-matrices over the $m$-indices yields the unit
matrix. 
When written in terms of the quantum group covariant basis,
Eq. (\ref{Liou}) simply becomes
the singlet formed from left and right moving spin $J$ representations
of highest and lowest weight, respectively. Remarkably, this mechanism
reproduces in particular the minimal models when one continues to central
charges $c<1$; that is, there is a description in terms of Liouville
exponentials of the observables of these models \cite{G93}. For $c>1$ (in fact,
$c>25$), the closest analog of the minimal models is Liouville theory 
restricted to the Kac table. The diagonal sums are finite (the $n$ -sum
in Eq. (\ref{Liou}) truncates at $n=2J$), and closure
of the chiral and non-chiral algebra become essentially equivalent.

When we leave the Kac table, the situation changes rather drastically: it is
no longer possible to give a {\it unique} representation of the Liouville
exponentials in terms of a sum over chiral vertex operators. As pointed
out in ref. \cite{Sch96}, this is due to two reasons: First, the formal 
expansion Eq. (\ref{Liou}) cannot be evaluated term by term, using naive
charge conservation rules, when the sum is infinite (see also ref. \cite{RPS98}
for a recent discussion on this fact from another point of view). Second,
when $2J$ is not an integer, the expansion Eq. (\ref{Liou}) suffers 
from a multi-valuedness
problem as regards its zero mode dependence, and therefore does not 
represent accurately the ``true'' Liouville exponentials even 
classically. According to the ideas of
ref. \cite{PRY}, adapted to the Liouville context in \cite{Sch96},
this does not mean that such expansions cannot be used;
rather, one needs to introduce a new (continuous) parameter $\beta$
which controls
the monodromy properties of the expansion with respect to the zero mode.
One arrives then at generalized expansions of the form
\beq
\sum_{n} a_{m;\beta}^{(J)}(\varpi) V^{(J)}_{-J}\Vb^{(J)}_{-J}
S^{\beta+n}\Sb^{\beta+n}.
\label{Liougen}
\eeq

For the three point function, with operators situated at $0,1,\infty$,
all $\beta$ parameters are fully determined, up to integers. For the
operator at $z=1$, not replaced by a highest weight state,
one has to choose $\beta$ such that the
total monodromy becomes trivial, i.e. the multi-valuedness
disappears. 
One can in particular consider the set of three point functions
where $\beta$ is of the form $n+\nhat{\alpha_+\over\alpha_-}$ ($n,\nhat$
integer), 
so that the exponential can be described by Eq. (\ref{Liou}) or its
generalization to both screenings $\alpha_+,\alpha_-$. This was
done in ref. \cite{Sch96}. In this context, one already needs the
negative screening operators discussed in the present paper, because
they are generated non-perturbatively by a careful evaluation of 
the expansion Eq. (\ref{Liou}). The non-perturbative contributions
can be rendered perturbative in terms of an ``effective'' representation
of the Liouville exponentials that can be evaluated using naive charge
conservation rules: 
\beq
e^{-J\alpha_-\Phi}_{\hbox{\tiny{eff}}}
=\sum_{n=-\infty}^\infty a_n^{(J)}(\varpi) U_m^{(J)}(u)\Ub_m^{(J)}(v)
\label{Lioueff}
\eeq
It is an interesting question whether Eq. (\ref{Lioueff}) 
is more than just a recipe for the computation of a class of
three point functions.
In ref. \cite{GS94} it was shown that Eq. (\ref{Liou}) is compatible with
locality in the sense that any two Liouville exponentials of the form 
(\ref{Liou}) will commute at equal times, order by order in powers
of screening operators (i.e. perturbatively). 
It was also observed that in the same sense,
the Liouville equation is fulfilled by the exponential with $J=-1$ (see also
\cite{OW86}). 

It is natural to ask whether the same properties will obtain 
non-perturbatively, i.e. when taking into account the additional negative
screening contributions of Eq. (\ref{Lioueff}). While this is essentially 
trivial for the equations of motion, non-perturbative locality requires
that 
\ben
 [e^{-J\alpha_-\Phi}(\tau,\sigma), \widetilde {e^{-J'\alpha_-\Phi}}
  (\tau,\sigma')]=0
\een
Here, both exponentials are given by Eq. (\ref{Liou}) and understood
to be treated perturbatively, but the second one
is constructed in terms of the free field $\tilde X$.
The results of the present paper provide all the necessary tools in order
to answer this question. One has to show that there is a new orthogonality
relation between $q$-deformed $6j$-symbols corresponding to positive and 
negative
screenings, respectively; work on this problem is in progress. 
If locality holds both in the perturbative and the non-perturbative sense,
at least for integer (positive or negative) $2J$, this would be a hint that 
there exist
consistent conformal field theories described by positive integer, and 
arbitrary integer screening powers, respectively. 

Some counter-evidence,
however, seems to be provided by the analysis of the ``perturbative''
operator product of two Liouville exponentials. While it formally closes
on the set of Liouville exponentials with integer positive screenings,
there appear arbitrarily strong short-distance singularities 
\cite{GS94}\footnote{The formula (\ref{fus5}) corrects a misprint in 
ref. \cite{GS94}. Note also that we are working here in Euclidean 
coordinates on the sphere where $z:=e^{\tau+i\sigma}$.}
that render the result dubious:
\bea
 &&e^{-J_1\alpha_-\Phi(z_1, \zb_1 )}
  e^{-J_2\alpha_-\Phi(z_2, \zb_2 )}=
  \sum_{ J_{12}=-\infty}^{J_1+J_2}  \sum _{\{\nu\}, \{\nub\} }
  e^{-J_{12} \alpha_-\Phi^{\{\nu\}, \{\nub\}}(z_2, \zb_2 )}\nn
 &\cdot&
  |z_1-z_2|^{2(\Delta_{J_{12}}-\Delta_{J_1}-\Delta_{J_2})}
  \langle\varpi_{J_{12}}, \varpib_{J_{12}}; \{\nu\}, \{\nub\} |
  e^{-J_1\alpha_-\Phi(z_1-z_2, \zb_1-\zb_2 )} |
  \varpi_{J_2}, \varpib_{J_2}\rangle
\label{fus5}
\eea
where $\varpi_J:=\varpi_0+2J\equiv 1+{\pi\over h}+2J$, and ${\nu}$ denotes
descendant contributions. The matrix element on the right hand side
represents the operator product coefficient as well as the descendant
contributions to the conformal block given by $J_{12}$; see ref. \cite{GS93}
for details. The contributions from very large negative $J_{12}$ are
non-vanishing in general and so there appear unwanted short-distance
singularities of arbitrarily high order as $\Delta_{J_{12}} \to -\infty$.

Even if there is a fully consistent non-chiral algebra involving
only integer screenings, we should be very careful
as to its interpretation as a restriction of Liouville theory. 
In fact, the operator product of Liouville
exponentials is expected \cite{ZZ96} to depend on all operators entering
in the correlator, and to involve ``hyperbolic'' states corresponding
to purely imaginary $\varpi$. Crossing symmetry, or locality, should then
involve the same intermediate channels, and the interpretation within
Liouville theory of locality
properties based on integer screenings only is a priori not very clear.
On the other hand, general $\beta$-dependent expansions of type
Eq. (\ref{Liougen}) would actually allow for the introduction of hyperbolic 
intermediate states. 
Consider a four point function with operators situated at $0,1,\infty$ 
and $z$, represented as
\ben
 \langle\varpi_4|e^{-J_3\alpha_-\Phi}(z,\bar z)e^{-J_2\alpha_-\Phi}(1)|
  \varpi_1\rangle
\een
The parameters $\beta_3,\beta_2$ appearing in the representation
of the exponentials in the middle are coupled by the monodromy condition.
For a given conformal block, they have a unique value. The question of
what range to choose for the one free $\beta$ parameter is thus tantamount
to the factorization problem, and imaginary $\beta$ will in fact produce
hyperbolic intermediate states. While in Liouville
theory restricted to the Kac table, locality and conformal invariance
are sufficient to fully determine the operator construction of the theory,
we see that in the present, much larger context this may well be an illusion.
The simple correspondence between chiral and non-chiral algebras seems to
be lost, and additional information is needed for the construction of 
non-chiral correlation functions.
A careful analysis of the zero mode dependence of the exponentials,
combined with
group-theoretical arguments as in refs. \cite{T97,T972}, seems to be a
step in the right direction.
The presence of a context-dependent, floating $\beta$ parameter as we
advocate it here,
means that general Liouville exponentials will invoke arbitrary quantum
group representations,
and not just highest or lowest weight ones as one could have naively
expected from the
classical picture. This would resemble the situation in affine $SL(2)$
current algebra where continuous representations of intertwining operators
appear for non-integrable weights \cite{PRY}. 
In any case, it seems clear that a deeper understanding of the full
Liouville dynamics
will have to pass through a study of the algebra of the chiral vertex
operators, and it seems desirable
to continue the program towards the analysis of arbitrary non-integer
screenings.
The study of correlation functions of the chiral vertex operators may be
interesting
in its own right, and is expected to lead to new mathematical functions
already on the
integer (including negative ones) screening level.

\subsection{Application to Strong Coupling Liouville Theory}

In a series of works \cite{GN85,G91,GR942+96,GR94},
Gervais and collaborators have proposed
a conformal field theory with $1<c<25$ which they interpret as a candidate
for Liouville theory in the forbidden, ``strong coupling'' region $1<c<25$.
This theory exists for certain discrete values 
of the central charge ($c=7,13,19$), and on
a spectrum of highest weight states specified by the unitary truncation
theorems of ref. \cite{G91}. Locality arises in a rather different way;
local observables become products of sums of chiral vertex operators
rather than sums of products, and each chiral factor is local up to a
phase factor. The local operators can be divided into two sets, one with
negative and one with positive conformal weights. 
The latter  operators involve negative screenings; they can be written
in the form (we write the left-moving factor only)
\beq
\chi_+^{(J)}(u)=\sum_{p\in \Nat_0} \, (-1)^{(2-s)p\left[2(p+{x-J\over 
{\pi\over h}-1})
+{p+1\over 2}\right]}
g^{x}_{J,x-J-1+p({\pi\over h}-1)} \ V^{(J)}_{p({\pi\over h}-1)-1-J}(u)
\label{chi}
\eeq
Here, $x:={1\over 2}(\varpi-\varpi_0)$ with $\varpi_0:=1+{\pi\over h}$,
and $s=0,\pm 1$ determines the central charge ($c=1+6(s+2)$). The operators
$V^{(J)}_m$ are nothing but normalized $U_m^{(J)}$ fields, 
$V^{(J)}_m\equiv {1\over I^{(J)}_m(\varpi)} U_m^{(J)}$. Finally, 
the coupling
constants $g^{J_{12}}_{J_1,J_2}$ appear as prefactors of the $6j$-symbols
in the fusion and braiding relations of the $V^{(J)}_m$ operators
and were determined in \cite{CGR941}. The values that $\varpi$ is allowed
to take are not arbitrary continuous, but discrete:
\ben
 \varpi=({\pi\over h}-1)(l+1+{r\over 2-s})
\een
with $l\in \Z$, $r=0,...,1-s$. Similarly, $J$ is restricted to be
\ben
 J=({\pi\over h}-1)(l'+{r'\over 1-s})-1
\een
with $l'\in \Z$, $r'=0,...,2-s$. 
The screening number with respect to $\alpha_-$
is $-p-1$, while for $\alpha_+$ it is $p$. Thus indeed, negative screenings
are involved, and one can obtain a second analogous set with the roles
of $\alpha_+$ and $\alpha_-$ interchanged.

In the absence of an algebraic control
over vertex operators with negative screenings, the properties of the $\chi$
fields were described
using the formal symmetry $J\to -J-1$ of the $q$-deformed
$6j$-symbols already mentioned above. 
The present analysis not only allows to  further corroborate 
this discussion (at least as far as the braiding is concerned) by 
establishing a direct derivation from the operator construction, 
but also opens the road towards an investigation of new
local operators with negative screenings which are 
not accessible by the $J\to -J-1$ analytic continuation technique.

The results of the present analysis and its possible future extensions
should also find direct applications in theories closely related to
Liouville,
such as $SL(2,\R)$ or $SL(2,\C)/SU(2)$ WZNW theory \cite{T97}.
{}From a larger perspective, this program can be viewed as a kind of
bootstrap approach
to irrational conformal field theory in the simplest context where the
chiral symmetry algebra
is just Virasoro. It is an interesting question whether all solutions
can be identified either
with (subsectors of) weak or strong coupling Liouville theory.
\\[.4cm]
{\bf Acknowledgment}\\[.2cm]
We would like to thank J. Teschner for useful discussions and comments. 
J.R. gratefully acknowledges the partial financial support from 
the Danish Natural Science Research Council, contract no. 9700517.
He also thanks Laboratoire de Math\'ematiques et Physique Th\'eorique,
Universit\'e de Tours and The Niels Bohr Institute, where parts of this paper
were written down, for their kind hospitality. 

\appendix
\section{Concatenation of Standard Transformations}

Here we shall present some techniques involving ST (\ref{ST})
relevant for the many manipulations of
$q$-deformed Saalschutzian ${}_4F_3$ hypergeometric functions employed in the
main body of the paper.

We are interested in the case where two of the four upper entries are
integers and one of the three lower entries is an integer. Since our
main focus shall be on the integer entries, let 
\ben
 \left(\begin{array}{c} A,\ \ B\\ C\end{array}\right)
\label{ABC}
\een
represent such a situation. The remaining 4 entries are generic only
subject to the requirement of the $q$-hypergeometric function
being balanced. In order to keep the 3 integer entry structure, there are
three classes of ST applicable to (\ref{ABC}). The first class
does not affect the three integers. In the language of (\ref{F}) and
(\ref{ST}), it amounts to 
\ben
 c=A,\ d=B,\ e=C
\label{c1}
\een
By construction, either $A$ or $B$ must be non-positive and here we have
assumed $B\leq0$. The second class leaves only $A$ and $B$ unchanged. 
Again we may assume that $B\leq0$ in which case the second class of
transformations is characterized by
\ben
 c=A,\ d=B,\ f=C
\label{c2}
\een
The third class only affects one of the two upper entries, and assuming
$B\leq0$ we have
\ben
 a=A,\ d=B,\ e=C
\label{c3}
\een

A simple inspection reveals that 48 different configurations
may be obtained by naive (and repeated) applications of ST. 
The lower entry may take on the 6 different values $C$, 
$1+A+B-C$, $1-A+B$ and 2 minus either of these. However, a priori
not all of the configurations thus obtained
are well defined since we must have $d\leq0$.
Nevertheless, a further inspection shows that all well defined 
configurations, the number of which depends on the relations
between $A$, $B$ and $C$, may be obtained from one another by well defined
concatenations only.

Let us illustrate the above by considering the well defined concatenation 
of STs transforming (\ref{R}) (subject to (\ref{selI})) into (\ref{RI>}).
We have
\bea
 \left(\begin{array}{c} -n_1,\ \ -n'\\ 1-n'+n_2\end{array}\right)&\rightarrow&
  \left(\begin{array}{c} -n_1,\ \ -n'\\ -n-n'\end{array}\right)\rightarrow
  \left(\begin{array}{c} -n_2,\ \ -n'\\ -n-n'\end{array}\right)\nn
 &\rightarrow&
  \left(\begin{array}{c} -n,\ \ -n_2\\ -n-n'\end{array}\right)\rightarrow
  \left(\begin{array}{c} -n,\ \ -n_2\\ 1+n'-n_2\end{array}\right)
\eea
Finally, a ST of the first kind may be needed in order to match
all the entries of the $q$-hypergeometric function. 
It is then straightforward to verify that multiplying
all prefactors encountered in this repeated use of (\ref{F}) with the
original ones in (\ref{R}), produces the ones in (\ref{RI>}).

An interesting property of ST is that (modulo extra ST) it commutes
with all well defined substitutions of the parameters 
\ben
 a,...,g\ \rightarrow\ a',...,g'=a'+b'+c'+d'-e'-f'+1
\een
keeping the
integer structure as in (\ref{ABC}). That is, $A'$, $B'$ and $C'$ are integers
and at least $A'$ or $B'$ is non-positive.
Based on our classification above, the proof is almost immediate.
Case 1 may be characterized by (\ref{c1}). For $B'\leq0$ the statement
is trivial whereas for $A'\leq0$ it becomes equivalent to
\ben
 \frac{\q{f-d}_{-c}\q{e+f-a-b}_{-c}}{\q{f}_{-c}\q{e+f-a-b-d}_{-c}}
  =\frac{\q{f-c}_{-d}\q{e+f-a-b}_{-d}}{\q{f}_{-d}\q{e+f-a-b-c}_{-d}}
\label{f1}
\een
which is readily verified. Case 2 may be characterized by (\ref{c2})
and again the statement is trivial for $B'\leq0$. For $A'\leq0$
it becomes equivalent to (\ref{f1}).
Case 3 may be characterized by (\ref{c3}) and the statement is
trivial for $B'\leq0$. For $A'\leq0$ we need an additional ST
in order to prove our assertion. Namely, employing ST twice
results in
\bea
 {}_4F_3\left(\begin{array}{llll}a,&b,&c,&d\\  e,&f,&g&{} \end{array}
 ;\ q,\ 1\right)
 &=&\frac{\q{f-c}_{-d}\q{e+f-a-b}_{-d}}{\q{f}_{-d}\q{e+f-a-b-c}_{-d}}\nn
 &\cdot&\frac{\q{b+c+d-e-f+1}_{-d}\q{a+d-f+1}_{-d}}{
  \q{c+d-f+1}_{-d}\q{a+b+d-e-f+1}_{-d}}\\
 &\cdot&{}_4F_3\left(\begin{array}{llll}a,&e-b,&e-c,&d\\  
  e,&e+f-b-c,&a+d-f+1&{} \end{array}
 ;\ q,\ 1\right)\nonumber
\eea
and the statement becomes equivalent to
\bea
 \frac{\q{f-d}_{-a}\q{e+f-b-c}_{-a}}{\q{f}_{-a}\q{e+f-b-c-d}_{-a}}&=&
 \frac{\q{f-c}_{-d}\q{e+f-a-b}_{-d}}{\q{f}_{-d}\q{e+f-a-b-c}_{-d}}\nn
 &\cdot&\frac{\q{b+c+d-e-f+1}_{-d}\q{a+d-f+1}_{-d}}{
  \q{c+d-f+1}_{-d}\q{a+b+d-e-f+1}_{-d}}
\eea
which is readily verified.

\section{Remarks on Uniqueness}

We start from the following simple recursion relation
\bea
 \delta_{l_2,n'}&=&e^{i\alpha_1-i\alpha_{0}}
  R(-1/2,J';\varpi-1)_{-1,n'}^{l_2-1,n'-l_2}
  R(1/2,J';\varpi)_{1,l_2-1}^{l_2, 0}\nn
 &+&R(-1/2,J';\varpi-1)_{-1,n'}^{l_2,n'-l_2-1}
  R(1/2,J';\varpi)_{1,l_2}^{l_2, 1}
\label{recrel}
\eea
where $e^{i\alpha_0}$ and $e^{i\alpha_1}$ are phase factors that will
be specified below. The $R$-matrices with spins $-1/2$ and $J'$ are
considered as unknowns, while the ones with spins $1/2$ and $J'$ are
known since the braiding of a degenerate field with any other primary
is determined by null vector decoupling equations \cite{GN84}. 
Eq. (\ref{recrel}), as a two-term recursion relation, has an infinity 
of solutions that can be labelled for instance by an initial
condition $R(-1/2,J';\varpi)_{-1,n'}^{l_{20}-1,n'-l_{20}}$. Note that while
we assume that $n'$ is a positive integer, we admit $l_2$-values
of the form $l_2=l_{20}+p$, with $l_{20}$ a fixed real number  and $p$
an arbitrary integer. We can trivially rewrite Eq. (\ref{recrel}) as
\ben
 \delta_{l_2,n'}= \sum_{n_2}\,  e^{i\alpha_1-i\alpha_{1+n_2-l_2}}
  R(-1/2,J';\varpi-1)_{-1,n'}^{n_2,n'-1-n_2}
  R(1/2,J';\varpi)_{1,n_2}^{l_2, 1+n_2-l_2}
\label{recrel'}
\een
since $1+n_2-l_2$ can take the values $0$ and $1$ only (degenerate
fields braid into degenerate fields). Now we multiply both sides
by $V^{(J')}_{-J'}S^{l_2}(\sigma')S^{n'-l_2}(\sigma)$ from the right and 
sum over $l_2=l_{20}, l_{20}\pm 1, l_{20}\pm 2,... $, or equivalently 
over $l_1\equiv 1+n_2-l_2$. When $l_{20}$ is integer, we get
\bea
 V^{(J')}_{-J'}S^{n'}(\sigma')&=&\sum_{n_2} \, e^{i\alpha_1-i\alpha_{l_1}}
  R(-1/2,J';\varpi-1)_{-1,n'}^{n_2,n'-1-n_2}\nn
 &\cdot&\sum_{l_1=0}^1 \, R(1/2,J';\varpi)_{1,n_2}^{1+n_2-l_1, l_1}\, 
  V^{(J')}_{-J'}S^{1+n_2-l_1}(\sigma')S^{n'-1-n_2+l_1}(\sigma),
\label{opform}
\eea
otherwise the left hand side vanishes. For definiteness of notation, we
will write down the equations for $l_{20}$ integer in the following. 
Let us now rewrite (somewhat artificially) $S^{n'-1-n_2+l_1}(\sigma)$ as 
\ben
 S^{n'-1-n_2+l_1}(\sigma)=V^{(1/2)}_{-1/2}S^{l_1}(\sigma)\odot
  V^{-(1/2)}_{1/2}S^{n'-1-n_2}(\sigma) e^{i\alpha_{l_1}}
\een
Here, the renormalized product of Eq. (\ref{odotdef}) reappears, and
the phase factor is given by Eq. (\ref{odot}), i.e. 
$e^{i\alpha_{l_1}}=q^{l_1}$. Thus, Eq. (\ref{opform}) becomes
\bea
 V^{(J')}_{-J'}S^{n'}(\sigma')&=&\sum_{n_2} \, e^{i\alpha_1}
  R(-1/2,J';\varpi-1)_{-1,n'}^{n_2,n'-1-n_2}
  \sum_{l_1=0}^{1}\, R(1/2,J';\varpi)_{1,n_2}^{1+n_2-l_1, l_1}\nn 
 &\cdot&V^{(J')}_{-J'}S^{1+n_2-l_1}(\sigma') V^{(1/2)}_{-1/2} S^{l_1}(\sigma)
  \odot V^{-(1/2)}_{1/2}S^{n'-1-n_2}(\sigma)
\eea
Here we recognize that the sum over $l_1$ in front of the renormalized product
is nothing but the result of the braiding of $V^{(1/2)}_{-1/2}S(\sigma)$ with
$V^{(J')}_{-J'}S^{n_2}(\sigma')$. Pulling out the first of these two
operators to the left, we have using the second line of Eq. (\ref{3.6}),
\bea
 V^{(J')}_{-J'}S^{n'}(\sigma')&=&\lim_{\sigma_1 \to \sigma}
\sum_{n_2} \, S V^{(1/2)}_{-1/2}(\sigma_1)
  \, R(-1/2,J';\varpi-1)_{-1,n'}^{n_2,n'-1-n_2}\nn
 &\cdot&V^{(J')}_{-J'}S^{n_2}(\sigma')V^{(-1/2)}_{1/2}S^{n'-1-n_2}(\sigma)
{1\over (1-e^{i(\sigma_1-\sigma)})^{-\Delta_{1/2}-\Delta_{-1/2}}}
\eea
Note that the remaining phase factor $e^{i\alpha_1}$ has been removed
by reordering $S(\sigma)$ and $V^{(1/2)}_{-1/2}(\sigma)$. Now we can remove
the factor $SV^{(1/2)}_{-1/2}(\sigma)$ on the right hand side by multiplying
both sides by $V^{(-1/2)}_{1/2}S^{-1}(\sigma)$, and we arrive finally at
\bea
 &&V^{(-1/2)}_{1/2}S^{-1}(\sigma) V^{(J')}_{-J'}S^{n'}(\sigma')\nn
 &=&\sum_{n_2}\,  R(-1/2,J';\varpi)_{-1,n'}^{n_2,n'-1-n_2}\ 
  V^{(J')}_{-J'}S^{n_2}(\sigma')V^{(-1/2)}_{1/2}S^{n'-1-n_2}(\sigma)
\label{initrel}
\eea
when $l_{20}$ is integer. This is the defining relation for the $R$-matrix  
$R(-1/2,J';\varpi)_{-1,n'}^{n_2,n'-1-n_2}$ and thus appears to be a 
triviality. However, in our derivation, $R(-1/2,J';\varpi)_{-1,n'}^{n_2,n_1}$ 
is, a priori, {\em not} an $R$-matrix but a solution to
Eq. (\ref{recrel}). In particular, it is {\em not unique}, as long 
as we do not impose our selection rule. So the surprising conclusion
is that there is an infinity of equivalent $R$-matrices for the braiding
problem Eq. (\ref{initrel}). This conclusion is not specific to the 
spin $J=-1/2$ and $n=-1$ but applies to type {\bf II} 
(and {\bf III},{\bf IV}) in general, 
as can be seen immediately by concatenation type 
arguments. When the starting value $l_{20}$ is not integer, the
left hand side of Eq. (\ref{initrel}) vanishes. This means that to any 
$R$-matrix with integer outgoing screening numbers, we can add
contributions with non-integer outgoing screenings, if the latter
are solutions to Eq. (\ref{recrel}). 

Of course, due to the presence of infinite sums one may question 
the rigorousness of various steps in the above argument, but it certainly
poses an interesting problem, which merits further investigation. 
Note that except possibly for the step from Eq. (\ref{recrel'}) to 
Eq. (\ref{opform}), the argument can also be read backwards. 

An interesting special case arises when 
$V^{(J')}_{-J'}S^{n'}(\sigma')$ is degenerate ($2J'$ positive integer, 
\ $ 0\le n'\le 2J'$) so that the null vector
decoupling equations control the braiding problem Eq. (\ref{initrel}). 
The resulting braiding matrix has  integer outgoing screenings
and is non-vanishing only for a finite number of $n_2$-values. 
This braiding matrix is the only solution to Eq. (\ref{recrel}) with
this property, and it is also the one that is produced by imposing
our selection rule - a welcome verification of the latter. 
However, Eq. (\ref{recrel}), even when
restricted to integer $l_2$, also allows for
solutions corresponding to infinite braiding sums in Eq. (\ref{initrel}),
with $n_2$ bounded from above but not from below. 
Indeed, the vanishing of $R(1/2,J';\varpi)_{1,l_2-1}^{l_2,0}$  
for $l_2=2J'+1$
implies that $R(-1/2,J';\varpi)_{-1,n'}^{2J'+1+i,n'-2J'-2-i} \, =0$ for
$i=0,1,2,...$ and $2J'$ positive integer, $ 0\le n'\le 2J'$. However,  
Eq. (\ref{recrel}) does not enforce any 
truncation of these matrix elements at large negative $i$, though
the equation at $l_2=n'$ {\it allows} for such a truncation to occur
through the vanishing of the $R$-matrix element $R(-1/2,J';\varpi)_{
-1,n'}^{l_2-1,n'-l_2}$ on the right hand side. 

{}From the point of view
of the monodromy analysis for solutions of the null vector decoupling
equations, the appearance of additional solutions seems to be
consistent only if
there exist formal infinite linear combinations of $q$-hypergeometric
functions (for $J'=1/2$) and their generalizations, with parameters
varying as a function of $n_2$, that vanish. The study of these questions
is left for further analysis.

\end{document}